# Doping dependent critical current properties in K, Co, and P-doped $BaFe_2As_2$ single crystals

Shigeyuki Ishida,[*] Dongjoon Song,[*] Hiraku Ogino, Akira Iyo, Hiroshi Eisaki
*Electronics and Photonics Research Institute, National Institute of Advanced Industrial Science and Technology (AIST), Tsukuba 305-8568, Japan*

Masamichi Nakajima
*Department of Physics, Osaka University, Toyonaka, Osaka 560-0043, Japan*

Jun-ichi Shimoyama
*Department of Physics and Mathematics, Aoyama Gakuin University, Sagamihara 252-5258, Japan*

Michael Eisterer
*Atominstitut, TU Wien, Stadionallee 2, 1020 Vienna, Austria*

[*] Both authors contributed equally to this work.

In order to establish the doping dependence of the critical current properties in the iron-based superconductors, the in-plane critical current density ($J_c$) of $BaFe_2As_2$-based superconductors, $Ba_{1-x}K_xFe_2As_2$ (K-Ba122), $Ba(Fe_{1-x}Co_x)_2As_2$ (Co-Ba122), and $BaFe_2(As_{1-x}P_x)_2$ (P-Ba122) in a wide range of doping concentration ($x$) was investigated by means of magnetization hysteresis loop (MHL) measurements on single crystal samples. Depending on the dopant elements and their concentration, $J_c$ exhibits a variety of magnetic-field ($H$)- and temperature ($T$)- dependences. (1) In the case of K-Ba122, the MHL of the under-doped samples ($x \leq 0.33$) exhibits the second magnetization peak (SMP), which sustains high $J_c$ at high $H$ and high $T$, exceeding $10^5$ A/cm$^2$ at $T = 25$ K and $\mu_0 H = 6$ T for $x = 0.30$. On the other hand, the SMP is missing in the optimally- ($x \sim 0.36$-$0.40$) and over-doped ($x \sim 0.50$) samples, and consequently $J_c$ rapidly decreases by more than one order of magnitude, although the change in $T_c$ is within a few K. (2) For Co-Ba122, the SMP is always present over the entire superconducting (SC) dome from the under- ($x \sim 0.05$) to the over-doped ($x \sim 0.12$) region. However, the magnitude of $J_c$ significantly changes with $x$, exhibiting a sharp maximum at $x \sim 0.057$, which is a slightly under-doped composition among Co-Ba122. (3) For P-Ba122, the highest $J_c$ is attained at $x = 0.30$ corresponding to the highest $T_c$ composition. For the over-doped samples, the MHL is characterized by a SMP located close to the irreversibility field $H_{irr}$. Common to the three doping variations, $J_c$ becomes highest at the under-doping side of the SC dome near the phase boundary between the SC phase and the antiferromagnetic/orthorhombic (AFO) phase. Also, the peak appears in a narrow range of doping, distinct from the $T_c$ dome with broad maximum. These similarities in the three cases indicate that the observed doping dependence of $J_c$ is intrinsic to the $BaFe_2As_2$-based superconductors. The scaling analysis of the normalized pinning force density $f_p$ as a function of the reduced magnetic field $h = H/H_{irr}$ ($H_{irr}$: irreversibility field) shows that the peak in the pinning force position ($h_{max}$) depends on $x$, indicating a change in pinning with $x$. On the other hand, high-$J_c$ samples always attain similar $h_{max}$ values of 0.40-0.45 for all the dopants, which may suggest that a common pinning source causes the highest $J_c$. A quantitative analysis of the $T$-dependent $J_c$ indicates that the two pinning mechanisms, namely, the spatial variations in $T_c$ (referred to as $\delta T_c$ pinning) and the fluctuations in the mean free path ($\delta l$ pinning), are enhanced for the under-doped samples, which results in the enhancement of $J_c$. Possible origins for the different pinning mechanism are discussed in connection with the $x$-dependence of $T_c$, the residual resistivity, AFO domain boundaries, and a possible quantum critical point.

## I. INTRODUCTION

The investigation of the critical current ($J_c$) properties is one of the main topics in the research of the iron (Fe) -based high-transition-temperature (high-$T_c$) superconductors. Fe-based superconductors are considered as promising candidates for large-current and/or high-magnetic-field applications, since they possess considerably high $T_c$ reaching 56 K at highest, as well as high upper critical magnetic fields ($H_{c2}$) exceeding 100 T with moderate anisotropy ($\gamma = H_{c2}^{ab}/H_{c2}^{c} = 1$-$10$) [1-3]. To evaluate their current carrying ability, various experiments have been carried out so far. In most cases, encouraging results have been obtained, such as a high $J_c$ exceeding $10^5$ A/cm$^2$ even under high-$H$s above 10 T [4,5], a moderate anisotropy of $J_c$ in superconducting tapes with respect to the direction of the applied $H$ [6,7], superior inter-grain connectivity [8], *etc*. It was also demonstrated that the introduction of artificial pinning centers by irradiation with heavy-irons, neutrons, electrons, *etc*. largely enhances $J_c$ [9-12]. At this moment, the improvement of inter-grain connectivity is one of the challenges for applications such as powder-in-tube wires and tapes [13,14]. At the same time, understanding of intra-grain $J_c$ properties and finding the way to enhance $J_c$ are also important because the enhancement of the intra-grain $J_c$ leads to an extension of the operation temperature and field range.

From a basic point of view, Fe-based superconductors are layered materials, as the cuprate high-$T_c$ superconductors. In the case of cuprates, vortices strongly interact with atomic-scale defects due to the short coherence length. The large magnetic penetration depth, the large anisotropy, as well as the high operation temperature amplify the effects of thermal fluctuations, and give rise to unprecedented phenomena, such as the elastic motion of the vortex lattice [15], plastic vortex motion [16,17], formation of a vortex glass [18], vortex melting [19], an order-disorder phase transition [20-22] of the vortex lattice or its structural phase transition [23], *etc*. One naturally expects that the vortex physics of Fe-based superconductors is as rich as in the cuprate counterparts.

So far, intrinsic $J_c$ properties of Fe-based superconductors and the relevant vortex physics have been investigated mainly on the following four systems, $Ln$FeAsO ($Ln$ = rare earth, 1111), $Ae$Ba$_2$As$_2$ ($Ae$ = alkali earth, 122), $A$FeAs ($A$ = alkali, 111), and FeSe (11) – based compounds. Here most of the experiments were carried out on single crystal samples, and $J_c$ was determined by magnetic hysteresis loop (MHL) measurements. To summarize the results, (1) $J_c$ of Fe-based superconductors is generally high: $J_c \sim 2$ MA/cm$^2$ at 5 K and 0 T for SmFeAsO$_{1-x}$F$_x$ [11], $\sim 2$ MA/cm$^2$ for K-Ba122 [24], $\sim 0.1$ MA/cm$^2$ for LiFeAs [25], and $\sim 0.2$ MA/cm$^2$ for FeTe$_{1-x}$Se$_x$ [26], respectively. (2) The MHL measurements show a sharp peak at around zero field and a power-law decay of $J_c$ at low-field region, which are attributed to the strong pinning [27,28]. (3) The second magnetization peak (SMP) in the MHL, characterized by a hump far below $H_{c2}$, is evident in most cases, which supports a high $J_c$ of Fe-based superconductors under high magnetic fields. As for the origin of SMP, (i) a crossover from elastic creep to plastic creep [29-34] or (ii) the corresponding order-disorder transition of the vortex lattice [35,36], (iii) a phase transition between a rhombic (at low $H$) and a high-$H$ square vortex lattice [25,37,38], *etc*. have been put forth. (4) Various source of pinning were proposed. Examples are : (i) dense vortex pinning nanostructures arising from inhomogeneous distributions of dopant atoms [39], (ii) charged dopant atoms [28], (iii) structural/magnetic domain boundaries [40,41], *etc*. (5) Two kinds of pinning mechanism, namely spatial variations in $T_c$ (referred to as δ$T_c$ pinning) and fluctuations in the mean free path (δ$l$ pinning) [42] are discussed based on the $T$-dependence of $J_c$ [30,33,37,43-45]. (6) As for the doping dependence of $J_c$, existing results are controversial. For example, (i) the doping dependence of $J_c$ follows the doping dependence of superconducting transition temperature ($T_c$) (for Co-Ba122 [30]), (ii) $J_c$ does not follow $T_c$ and becomes highest at the under-doped region (for Co-Ba122 [40] and K-Ba122 [46]), and (iii) $J_c$ is determined by some extrinsic factors rather than by the doping concentration (for P-Ba122 [47]). At this moment consensus has not been settled yet.

In order to understand the pinning mechanism of Fe-based superconductors and to find a way to enhance their $J_c$, the establishment of doping dependence of $J_c$ is useful, which enables us to discuss the relationship between $J_c$ and system specific properties, such as $T_c$, $H_{c2}$, anisotropy, dopant element, *etc*. However, existing experimental results suggest that $J_c$ of Fe-based superconductors also depends on extrinsic factors such as the quality of samples. Therefore, to deduce the inherent trends in behavior, systematic studies on various kinds of well-characterized samples are indispensable. For this purpose, 122 materials are most suitable because of a number of reasons. First, high-quality single crystals with various doping elements (K for Ba, Co for Fe, and P for As) and with different doping concentration ($x$) are available. Second, 122-based compounds are the most promising candidates for practical applications among the known Fe-based superconductors [1-3] and any knowledge for enhancing $J_c$ is therefore meaningful and valuable. This is one of the reasons why the pinning properties of 122 materials are investigated intensively.

In this study, we carried out systematic experiments on the critical current properties in K-, Co-, and P-Ba122 using high-quality single crystals. This is a comparative study that overviews the doping dependence of the critical current properties in these three materials leading to a comprehensive understanding of the pinning mechanism in Fe-based superconductors. By examining a large number of samples, we successfully established the doping dependence of $J_c$ for the three cases. We demonstrated $J_c$ and the SMP effect significantly depend on the dopant elements and their concentrations. On the other hand, we found that $J_c$ is largest for all dopants at the slightly under-doped composition, near the phase boundary between the SC phase and the antiferromagnetic-orthorhombic (AFO) phase. The sharp peak of $J_c$ at a particular doping concentration is distinct from the $T_c$ dome with broad maximum. The scaling analysis of the normalized pinning force ($f_p(h)$) density shows that the peak position of $f_p(h)$ ($h_{max}$) takes a similar value of 0.40-0.45 for the highest-$J_c$ samples, suggestive of a common mechanism which yields high $J_c$. The analysis of the $T$-dependence of $J_c$ indicates that the strength of both pinning mechanisms, δ$T_c$ pinning and δ$l$ pinning, is significantly larger in under-doped than in optimally or over-doped samples. These similarities found in the three cases indicate the presence of a common pinning source. Possible reasons for this enhanced pinning are discussed in detail.

The paper is organized as follows. In Sec. II, we describe the experimental procedures. In Sec. III, we show the $T$- and

$H$- dependence of $J_c$ for K-Ba122 (Sec. III A), Co-Ba122 (Sec. III B), and P-Ba122 (Sec. III C) with various $x$s obtained from MHL measurements. Related physical properties, such as in-plane resistivity ($\rho_{ab}(T)$) and $H_{c2}(T)$, are also presented. In Sec. IV, the $H$-dependence of the pinning force density and the $T$-dependence of $J_c$ are analyzed, and the possible source of pinning is discussed. In particular, we propose that $\delta l$ pinning is responsible for the significantly enhanced $J_c$ in highest-$J_c$ K-Ba122 crystal. Conclusions are drawn in Sec. V.

## II. EXPERIMENT

Single crystals of Ba$_{1-x}$K$_x$Fe$_2$As$_2$, Ba(Fe$_{1-x}$Co$_x$)$_2$As$_2$, and BaFe$_2$(As$_{1-x}$P$_x$)$_2$ were grown by the flux method using self-flux (KAs, FeAs, and Ba$_2$As$_3$/Ba$_2$P$_3$, respectively) following Ref. [48-50]. The compositions of the single crystals were determined by energy-dispersive X-ray (EDX) analysis and the X-ray diffraction using Cu K$\alpha$ radiation. The $c$-axis lengths determined by X-ray diffraction were consistent with the compositions obtained from EDX. The samples were cut into rectangular shapes with typical dimensions of ~1-2 mm (length) × 0.5-1 mm (width) × 0.02-0.1 mm (thickness) for the magnetization ($\chi(T)$) and the in-plane resistivity ($\rho_{ab}(T)$) measurements. The $T$- and $H$-dependences of the magnetization ($M(T, H)$) for $H // c$ were measured using a magnetic property measurement system (MPMS, Quantum Design). $\rho(T)$ measurements were carried out by a standard four probe method using a physical property measurement system (PPMS, Quantum Design). For the estimation of $J_c$, we applied the Bean model [51], i.e.

$$J_c = 20\Delta M/[w(1-w/3l)]$$

where $\Delta M$ is the width of the magnetization hysteresis loop (MHL) in the unit of emu/cm$^3$, $l$ is the sample length, and $w$ is the sample width ($l > w$).

## III. RESULTS

### 1. K-Ba122

#### 1. Sample characterization

Fig. 1(a) shows $\chi(T)$ for the K-Ba122 single crystals with their doping levels ranging from the under- ($x = 0.23$) to the over-doped ($x = 0.50$) region. To make the comparison easier, the data are normalized using the 5 K values. A sharp superconducting transition with a transition width ($\Delta T_c$) of 0.5-1 K was observed except for $x = 0.23$ and 0.25 ($\Delta T_c$ ~ 2-3 K). $T_c$ was defined as the midpoint of the transition and plotted as a function of $x$ in Fig. 1(i). $T_c$ increases with $x$ up to 38 K for $x = 0.36$ and decreases upon further doping down to 34 K at $x = 0.51$, thus forming a SC dome. The slope of the SC dome, $dT_c/dx$ is ~ 1.5 K per percent K on the under-doped side and -0.5 K per percent on the over-doped side. The broader transition of the under-doped samples might be associated with the larger slope, which causes a larger $\Delta T_c$ due to the spatial variation in $x$.

Fig. 1(b) shows $\rho_{ab}(T)$ for K-Ba122 (upper panel) and their differential curves d$\rho_{ab}$/d$T$ (lower panel) below 150 K. For $x$ = 0.23 and 0.25, d$\rho_{ab}$/d$T$ exhibits kink features at $T$ ~ 90 K and 65 K as indicated by arrows, which correspond to the magnetostructural phase transition from the high-$T$ paramagnetic/tetragonal (PT) phase to the low-$T$ antiferromagnetic/orthorhombic (AFO) phase [52]. The kink behavior is absent for $x = 0.29$ and above. The phase transition temperatures, $T_{s/N}$, are also plotted in Fig. 1(i).

Figs. 1(c)-(h) show $\rho_{ab}(T)$ of the $x = 0.23, 0.25, 0.29, 0.33, 0.36,$ and 0.51 samples under magnetic fields $\mu_0 H$ = 0-9 T applied parallel to the $c$-axis (upper panels) and to the $ab$-plane (middle panels). The resistive transition at 0 T is sharp with $\Delta T_c$ ~ 0.5-1.0 K except for $x = 0.23$ ($\Delta T_c$ ~ 2 K), and it shifts towards lower $T$ at higher fields. Small, but finite broadening is recognized with increasing $H$, particularly for $H // c$. The upper critical fields along the $c$-axis $H_{c2}^c(T)$ (squares) and along the $ab$-plane $H_{c2}^{ab}$ (triangles), defined by 90 % of the normal-state resistivity, are plotted in the bottom panel. The slope (d$H_{c2}$/d$T$) was determined using data at $\mu_0 H$ ≥ 1 T where $H_{c2}(T)$ is practically linear in $T$. The slope becomes highest at $x$ = 0.3-0.4, with d$H_{c2}^{ab}$/d$T$ ~ 12 T/K and d$H_{c2}^c$/d$T$ ~ 6.3 T/K, respectively. For $x = 0.23$, the resistive transition occurs above $T_c$ determined from $\chi(T)$, presumably due to the presence of higher-$T_c$ segments arising from the spatial inhomogeneity of dopant elements. $H_{c2}$ at 0 K, $H_{c2}(0)$, was estimated by the extrapolation of the linear fit to $H_{c2}(T)$ and the Werthamer-Helfand-Hohenberg (WHH) formula, $H_{c2}(0) = -0.69T_c(dH_{c2}/dT)$ [53]. The results are plotted in Fig. 1(j). The grey symbols indicate that the value may be overestimated owing to the broad resistive transition. Corresponding to the maximum $T_c$ and the maximum d$H_{c2}$/d$T$ slope, $H_{c2}(0)$ takes the maximum value of ~170 T ($H // c$) and ~300 T ($H // ab$) for $x = 0.3$-0.4, and rapidly decreases both on the under- and the over-doped sides. Fig. 1(k) shows the anisotropy factor of $H_{c2}$, $\gamma = (dH_{c2}^{ab}/dT) / (dH_{c2}^c/dT)$, as a function of $x$. In the investigated temperature range, $\gamma$ takes 1.6-2.2 and does not show a noticeable $x$-dependence. It should be noted that the WHH formula is valid for a superconductor in the clean limit with a single-band spherical Fermi surface and the obtained $H_{c2}(0)$ values are not quantitatively correct (in the present case, $H_{c2}^{ab}(0)$ are overestimated). In any case, $H_{c2}(0)$ values of present work are in good agreement with previous results which have been also estimated by WHH method [54]. This indicates that the quality of the present samples is good and resistivity measurement is properly done.

#### 2. Magnetization hysteresis loops

The magnetization hysteresis loops (MHLs) for $x = 0.23, 0.25, 0.29, 0.33, 0.36,$ and 0.51 samples are shown in Figs. 2(a)-(f) (5 K ≤ $T$ ≤ 0.7$T_c$) and Figs. 2(g)-(l) ($T$ ≥ 0.7$T_c$), respectively. The overall behavior of the MHL changes with $x$, which are classified into three groups, namely, Group (1): under-doped samples ($x = 0.23$, 0.25, and 0.30), Group (2): slightly under-doped sample ($x = 0.33$), and Group (3): optimally- to over-doped samples ($x = 0.36$ to $x = 0.51$). Note that the shape of the MHL dramatically changes between $x = 0.30$ and $x = 0.36$, whereas $T_c$ differs only by 1.5 K. In the following, the characteristics of the MHLs in these three groups are compared.

Group (1): under-doped samples ($x = 0.23$, 0.25, and 0.30): The MHL at $x = 0.23$ (Fig. 2(a)) is symmetric with respect to both $H$- and $M$-axes. This fact indicates that the dominant pinning mechanism is of bulk nature, and thus justifies the application of the Bean model for the estimation of $J_c$. The $H$-dependence of $M$ is non-monotonic. Following a sharp peak near $H = 0$, $M(H)$ decreases and then increases with $H$. The behavior is associated with the second magnetization peak (SMP), which enters into the measurement range ($\mu_0 H < 7$ T) above 10 K. The peak position, defined as $H_{sp}$, moves towards lower $H$ with increasing $T$ and persists up to $T = 20$ K (Fig. 2(g)). The MHLs for $x = 0.25$ (Figs. 2(b) and (h)) and 0.30 (Figs. 2 (c) and (i)) are qualitatively similar to the one for $x = 0.23$, while the magnitude of $M$ and $H_{sp}$ are larger. The results are consistent with those reported by Yang *et al.* [24] and Kim *et al.* [55]

Group (2): slightly under-doped sample ($x = 0.33$): As shown in Figs. 2(d) and (j), the SMP effect is also present for $x = 0.33$. While the behavior is apparently similar to that in Group (1) at low $T$, the MHLs at different $T$ cross each other at high $T$ (Fig. 2(j)). This behavior implies that the $M(T)$ curve at fixed $H$ exhibits a non-monotonous $T$-dependence, which possesses a peak at high $T$. Such a behavior does not occur in the Group (1) samples, in which $M$ monotonically decreases with $T$ at any $H$. Moreover, the magnitude of $M$ decreases by nearly one order of magnitude compared to $x = 0.30$, even though their $T_c$s are comparable. A similar SMP behavior is also observed in over-doped P-Ba122, which will be shown later.

Group (3): optimally to over-doped samples ($x = 0.36$ to $x = 0.51$): Here the width of the MHLs monotonously decreases with increasing $H$, in other words, the SMP effect is missing. See Figs. 2(e), (f), (k), and (l). As a consequence, the magnitude of $M$ is small, particularly at high $H$. The disappearance of the SMP effect is unique to K-Ba122, not observed in Co- or P-Ba122 at any doping levels.

The present results indicate that the MHLs of K-Ba122 significantly depend on $x$. So far, some groups reported the existence of the SMP effect in optimally-doped K-Ba122 [24,31] whereas other groups reported its absence [44,56]. The apparent discrepancy likely comes from a slight difference in the effective $x$.

### 3. *Critical current density and vortex phase diagram*

Based on the MHLs, we calculated $J_c$ by employing the Bean model [51]. The results are shown in Figs. 2(m)-(r). Here, the $y$-scales of these Figures differ from each other. To make the comparison easier, red dashed straight lines mark $J_c = 0.1$ MA/cm$^2$ in the Figures.

The Group (1) samples, $x = 0.23$ (Fig. 2(m)), $x = 0.25$ (Fig. 2(n)), and $x = 0.30$ (Fig. 2(o)), generally possess large $J_c$ of the order of $0.1 – 1$ MA/cm$^2$ over a wide $T$- and $H$- range. Among them, the largest $J_c$ reaches 2.5 MA/cm$^2$ at $T = 5$ K, $\mu_0 H = 0$ T for the $x = 0.30$ sample. The $H$-dependence of $J_c$ is rather weak in these samples, which is due to the SMP effect existing at finite $H$.

$J_c$ decreases for the Group (2) ($x = 0.33$, Fig. 2(p)) sample. The $J_c$ of 1.2 MA/cm$^2$ at $T = 5$ K and $\mu_0 H = 0$ T is about half the one at $x = 0.30$. $J_c = 0.01$ MA/cm$^2$ at $T = 30$ K and $\mu_0 H = 2$ T is one order of magnitude smaller than $J_c$ for $x = 0.30$ under the same conditions, even though their $T_c$s are almost the same.

For the Group (3) samples, $x = 0.36$ (Fig. 2(q)) and $x = 0.51$ (Fig. 2(r)), $J_c$ monotonously decreases with $H$ and its values become smaller, particularly at high $T$ and high $H$. Above 20 K, $J_c$ of the $x = 0.36$ sample is more than one order of magnitude smaller compared to $x = 0.30$. This result is rather unexpected since the crystal with $x = 0.36$ has the highest $T_c$ among the K-Ba122 samples and is thus regarded as the optimal composition. For $x = 0.51$, $J_c$ decreases by another order of magnitude, down to 0.001 MA/cm$^2$ at 20 K and 6 T, despite its $T_c$ is still as high as 33 K. The co-occurrence of high-$T_c$ and low-$J_c$ in Group (3) contrasts with low-$T_c$ and high-$J_c$ in the Group (1) counterpart.

Figs. 2(s)-(x) show the $T$- and $H$- dependence of $J_c$ in form of contour plots. In the Figures, several characteristic magnetic fields are also marked; $H_{on}$: the onset of the SMP effect defined by the local minimum of the MHL (pink circle), $H_{sp}$: the SMP peak position (red diamond), $H_{irr}$: irreversibility field defined by a criterion of $J_c < 100$ A/cm$^2$ (yellow square), and $H_{c2}{}^c$: upper critical field along the $c$-axis obtained from the resistivity measurements (orange triangle). These characteristic magnetic fields divide the superconducting state into four regions, namely, Region (I): below $H_{on}(T)$, Region (II): between $H_{on}(T) – H_{sp}(T)$, Region (III): between $H_{sp}(T) – H_{irr}(T)$, and Region (IV): between $H_{irr}(T) – H_{c2}(T)$.

For the Group (1) samples, $x = 0.23$ (Fig. 2(s)), $x = 0.25$ (Fig. 2(t)) and $x=0.30$ (Fig. 2(u)), high-$J_c$ areas, which are coded by bright colors, extend over a wide $T$- and $H$- range. Furthermore, it is noticed that the red- and yellow-colored areas in Fig. 2 (t) and the green- and light blue-colored areas in Fig. 2(u) fan out with increasing $H$. This behavior reflects the SMP effect. With increasing $x$, $H_{sp}$ increases, resulting in the expansion of Region (II). In particular, for $x = 0.30$, Region (II) dominates the superconducting phase, due to large $H_{sp}$ even at high $T$. For $x = 0.25$ and $x = 0.30$, $H_{irr}$ exists close to $H_{c2}$. This is consistent with the small broadening of $\rho_{ab}(T)$ and implies a weak thermal fluctuation, which is a favorable property for potential applications requiring a large current carrying ability. $H_{c2}(T)$ is not plotted for $x =$

0.23 (Fig. 2(s)), due to the uncertainty in determining $H_{c2}(T)$ from $\rho_{ab}(T)$ for this sample.

In the case of $x = 0.33$ (Fig. 2(v)), which is classified into Group (2), the phase diagram is also divided into four regions reflecting the SMP effect. Here Region (I) occupies a larger area and high $J_c$ is realized within a limited space in Region (I), only at low $T$ and low $H$. Indeed in this case, Region (II) is colored in blue, corresponding to $J_c < 0.1$ MA/cm$^2$. The facts indicate that the SMP does not lead to high $J_c$ for $x = 0.33$, in contrast to Group (I).

For $x = 0.36$ (Fig. 2(w)) and $x = 0.51$ (Fig. 2(x)), the phase diagrams are separated into two regions, i.e., Region (I) and (IV), due to the absence of the SMP. A high $J_c$ is observed only at low $T$- and low $H$-regions as in the case of $x = 0.33$. For $x = 0.36$, $H_{irr}(T)$ is smaller compared with the $x = 0.30$ and 0.33 samples, while $H_{c2}(T)$ is almost identical for all compositions, leading to an expansion of Region (IV). This expansion is also recognized for $x = 0.51$. The small $H_{irr}(T)$ would suggest weak pinning, which will be discussed later.

The contour plots shown in Figs. 2(s)-(x) are regarded as the vortex phase diagrams of K-Ba122. In general, the vortex phase diagram is determined by the competition of the elastic energy of vortex lattice, the pinning energy, and the thermal energy. The SMP effect is commonly associated with an order-disorder transition, which occurs when the pinning energy exceeds the elastic energy. In previous studies of the 122-type Fe-based superconductors, two scenarios were proposed in discussing the vortex phase diagram associated with the SMP effect, namely, (i) the SMP is due to the crossover from the elastic creep to plastic creep [29-31] and (ii) SMP comes from the structural phase transition of the vortex lattice [37,38]. Based on the scenario (i), in Region (I) and (II), the motion of vortex is governed by the elastic creep, while it is governed by the plastic creep in Region (III). In this case, $H_{sp}$ corresponds to the threshold field of the elastic-plastic crossover. Note that this scenario is compatible with the idea of an order-disorder transition, where $H_{on}$ roughly corresponds to the onset of the transition. On the other hand, based on the scenario (ii), $H_{sp}$ corresponds to the phase transition field of the vortex lattice from the rhombic to the square structure. In Region (II), the vortex lattice softens as $H$ approaches to $H_{sp}$, and the vortices are pinned more easily, resulting in increasing $J_c$ [57]. In the following we briefly examine whether and how one can explain the present results based on the above scenarios.

In general, $J_c$ depends on the combination of two factors, namely, the elementary pinning force and the density of pinning centers. The pinning force is related to the condensation energy of a superconductor, since it determines the energy gain when the vortex core is located on the pinning centers. Since $T_c$ and $H_{c2}$ increases from $x = 0.23$ to 0.30, the increase in $J_c$ in the Group (1) samples can be attributed to the increase in the condensation energy. Here, both two scenarios account for the observed $x$-dependence. On the other hand, $J_c$ decreases for $x = 0.33$ and $x = 0.36$ in spite of the increasing $T_c$ and $H_{c2}$, which is unlikely to result for a decreasing condensation energy. One needs to assume that the density or the strength of the pinning centers significantly decrease from $x = 0.30$ to 0.36. The model of the order-disorder transition predicts that $H_{on}$ and $H_{sp}$ move closer to $H_{irr}$ in this case [58], as experimentally observed. The crossing of MHLs at different $T$ is often observed for comparatively small pinning when the peak position is already close to $H_{irr}$, as seen in $x = 0.33$ (Group (2)). If the pinning energy does not exceed the elastic energy in the entire field range, the SMP disappears owing to the absence of the order-disorder transition as in the case for $x \geq 0.36$ (Group (3)). Thus, the doping evolution of vortex phase diagram from Group (1) to (3), in particular the monotonous increase of $H_{on}$ with $x$, can be understood based on the order-disorder transition scenario by assuming a continuous weakening of pinning with increasing dopant concentration, which is consistent with the observed decrease in resistivity. Weak pinning also accounts for the lower $H_{irr}$ (the expansion of Region (IV)) for $x \geq 0.36$. The $x$-dependence of the pinning behavior will be discussed quantitatively in the Discussion Section.

### 4. Doping dependence of critical current density

Figs. 3(a)-(c) show $J_c(H)$ for $x = 0.23$-0.51 at $T = 5$ K, $T = 0.5T_c$ and $0.8T_c$. Based on these results, we construct the contour plots of $J_c(x, H)$ in Figs. 3(d)-(f). One can immediately find that the behavior in the under-doped ($x < 0.33$) region is distinct from the optimally- and the over-doped ($x > 0.33$) counterpart. Below $x = 0.33$, high $J_c$ is sustained up to $\mu_0 H = 7$ T. Above $x = 0.33$, $J_c$ decreases abruptly, especially in magnetic fields.

At $T = 0.5T_c$, (Fig. 3(e)), a local maximum in $J_c$ occurs at 6 T for $x = 0.25$. This is due to the SMP effect. At $T = 0.8T_c$ (Fig. 3(f)), the composition with the highest $J_c$ depends on $H$. Below $\mu_0 H = 2$ T, the highest $J_c$ is obtained at $x = 0.25$, while it shifts to $x = 0.30$ at higher fields $\mu_0 H > 3$ T.

## 2. Co-Ba122

### 1. Sample characterization

Fig. 4(a) shows the normalized $\chi(T)$ for Co-Ba122 single crystals with $x = 0.05$, 0.057, 0.06, 0.08, 0.10, and 0.12. A sharp superconducting transition with $\Delta T_c \sim 1$ K was observed except for the under-doped ($x = 0.05$) and the heavily over-doped ($x = 0.12$) samples ($\Delta T_c \sim 2$-3 K). In Fig. 4(i), the $x$-dependence of $T_c$ is plotted. $T_c$ increases with $x$ up to $x = 0.06$ ($T_c = 24$ K) and decreases with further doping, down to 11 K at $x = 0.12$. $dT_c/dx$ is ~ 12 K per percent Co on the under-doped side and ~ -2 K on the over-doped side. The magnitude of the slope is larger compared with K-Ba122 and P-Ba122 (shown later) on both sides.

Fig. 4(b) shows $\rho_{ab}(T)$ (upper panel) and $d\rho_{ab}/dT$ (lower panel) below 150 K. For $x = 0.05$, $\rho_{ab}(T)$ shows an upturn at low $T$, which produces a shoulder and a dip in $d\rho_{ab}/dT$, at $T$

~ 65 K and 55 K, respectively, as indicated by black arrows. They correspond to the structural ($T_s$) and the magnetic ($T_N$) phase transition temperatures, respectively [59,60]. For $x = 0.057$, the features in $d\rho_{ab}/dT$ are not clear. We defined $T_s$ as the temperature where the slope of $d\rho_{ab}/dT$ is largest, indicated by a blue arrow. For $x = 0.06$ and above, $\rho_{ab}(T)$ and $d\rho_{ab}/dT$ do not show an appreciable feature, suggesting that there is no phase transition. $T_s$ and $T_N$ are also plotted in Fig. 4(i).

Figs. 4(c)-(h) show $\rho_{ab}(T)$ of Co-Ba122 with $x = 0.05$-$0.12$ for $H // c$ (top panels) and $H // ab$ (middle panels). As in the case of K-Ba122, the resistive transition shifts towards low $T$ with increasing $H$ with a weak broadening. In the bottom panels, $H_{c2}^c$ (solid circles) and $H_{c2}^{ab}$ (open circles) are plotted. $H_{c2}(0)$ was estimated using the WHH formula and its $x$-dependence is plotted in Fig. 4(j). $H_{c2}(0)$ takes maximum values of 60 T for $H // c$ and 150 T for $H // ab$, respectively at $x = 0.06$, and rapidly decreases along the under- ($x = 0.05$) and the over- ($x = 0.12$) doped sides. For $x = 0.05$ and 0.12, $T_c$ defined by $\rho_{ab}(T)$ is higher than $T_c$ defined by $\chi(T)$, thus the obtained $H_{c2}(0)$ values are overestimated ones. The data of the two samples are indicated by grey symbols. As shown in Fig. 4(k), $\gamma$ takes values in the range of 1.7-2.7 and does not show a noticeable $x$ dependence, similar to the case of K-Ba122.

### 2. Magnetization hysteresis loops

Figs. 5(a)-(l) show the MHLs for Co-Ba122 with $x = 0.05$-$0.12$ at selected $T$s (Figs. 5(a)-(f): $5\ K \leq T \leq 0.7T_c$, Figs. 5(g)-(l): $T \geq 0.7T_c$). In contrast to K-Ba122, all the MHL curves possess SMPs. For the most under-doped sample, $x = 0.05$ (Figs. 5(a) and (g)), the MHL shows a symmetric shape with a sharp central peak. At 5 K (black curve in Fig. 5(a)), the SMP is located at $\mu_0 H_{sp} = 1.2$ T, and moves to lower $H$ with increasing $T$, finally merges to zero field at $T \sim 8$ K. With increasing $x$, the SMP becomes evident, located at higher $H$. For example, for the optimally-doped sample, $x = 0.06$ (Figs. 5(c) and (i)), $H_{sp}$ is located above 7 T at 5 K. With increasing $T$, $H_{sp}$ moves to lower $H$, while it persists up to 22 K (~ $0.9T_c$). For the over-doped samples, $x = 0.08$-$0.12$, the SMP feature is still evident (Figs. 5(d)-(f) and (j)-(l)). $H_{sp}$ decreases monotonously with increasing $x$, from $\mu_0 H_{sp} > 7$ T for $x = 0.08$ to $\mu_0 H_{sp} \sim 1.4$ T for $x = 0.12$ at 5 K. Correspondingly, the magnitude of $M$ decreases with $x$.

### 3. Critical current density and vortex phase diagram

Figs. 5(m)-(q) show the calculated $J_c$ at selected $T$s for $x = 0.05$-$0.12$. The red dashed lines correspond to $J_c = 0.1$ MA/cm$^2$. $J_c$ shows a similar $H$- and $T$-dependence, while the magnitude strongly depends on $x$. The largest $J_c = 1.2$ MA/cm$^2$ at $T = 5$ K and $\mu_0 H = 0$ T was recorded for the slightly under-doped ($T_c = 22.5$ K) sample $x = 0.057$ (Fig. 5(n)). In this sample, $J_c$ exceeds 0.1 MA/cm$^2$ even under a high magnetic field of $\mu_0 H = 6$ T and up to a high temperature $T = 14$ K. A similarly large $J_c = 1.0$ MA/cm$^2$ ($T = 5$ K and $\mu_0 H = 0$ T) was also obtained for the optimally-doped ($T_c = 24.0$ K) sample with $x = 0.06$ (Fig. 5(o)). On the other hand, $J_c$ of the over-doped samples significantly decreases. For $x = 0.08$ (Fig. 5(p)), $J_c$ declines to 0.2 MA/cm$^2$ at $T = 5$ K and $\mu_0 H = 0$ T, which is one fifth of the value for $x = 0.06$, while $T_c = 20.5$ K is relatively high. With further doping, $J_c$ decreases to 0.06 MA/cm$^2$ for $x = 0.12$, which is one order of magnitude smaller than for the under-doped sample $x = 0.05$, even though $T_c$ of these two samples are comparable. A substantial degradation in $J_c$ in the over-doped region is common to K- and Co-Ba122.

Figs. 5(s)-(x) display the $T$- and $H$-dependence of $J_c$ by means of contour plots. Here the color scale changes from $x \leq 0.06$ to $x \geq 0.08$. We also plotted the characteristic magnetic fields, $H_{on}$ (pink circle), $H_{sp}$ (red diamond), $H_{irr}$ (yellow square), and $H_{c2}$ (orange triangle). Because the SMP effect is seen in all the samples, the $H - T$ contour maps are always separated into Region (I) to (IV).

For $x = 0.05$ (Fig. 5(s)), $H_{on}$ is located below 1 T, and $H_{sp}$ is about 0.5 T larger than $H_{on}$. Region (I) and (II) cover a small area at low $T$ and $H$. In this Figure, $H_{c2}(T)$ is not plotted due to the uncertainty in the definition of $H_{c2}$ from the resistivity data.

With $x$ increasing to 0.057 (Fig. 5(t)), the high current area expands to higher $T$ and $H$. Also, $H_{on}$ and $H_{sp}$ increase compared with $x = 0.05$, resulting in the expansion of Region (I) and (II). $H_{irr}(T)$ is close and nearly parallel to $H_{c2}(T)$, reflecting the small resistive broadening. For $x = 0.06$ (Fig. 5(u)), $H_{on}$ is similar to, and $H_{sp}$ is larger than for $x = 0.057$. As a result, Region (II) is expanded and the high $J_c$ area extends up to high $H$ along the $H_{sp}$ line.

For $x = 0.08$ (Fig. 5(v)), the high $J_c$ region shrinks and the light-blue color region becomes dominant, corresponding to an abrupt decrease in $J_c$. On the other hand, $H_{on}$ and $H_{sp}$ do not show an appreciable difference compared to $x = 0.06$. With further doping to $x = 0.10$ (Fig. 5(w)) and $x = 0.12$ (Fig. 5(x)), the light-blue region shrinks reflecting a further decreases in $J_c$. Also, both $H_{on}$ and $H_{sp}$ become smaller (Region (I) and (II) shrink). For $x = 0.08$, $H_{c2}(T)$ is not plotted for the same reason as for $x = 0.05$.

### 4. Doping dependence of critical current density

Figs. 6(a)-(c) show $J_c (H)$ of Co-Ba122 with various $x$s at $T = 5$ K, $0.5T_c$, and $0.8T_c$. The $x$-dependence of $J_c$ is visualized in Figs. 6(d)-(f) by means of a contour plot. At 5 K (Fig. 6(d)), $J_c$ rapidly increases with $x$ and attains the maximum value at $x = 0.057$. Around $x = 0.057$-$0.06$, high $J_c$ is sustained up to 7 T. Then $J_c$ rapidly decreases toward $x = 0.08$. Light-blue area extends at finite $H$ in the over-doped region, which reflects the non-monotonic $H$-dependence of $J_c$ due to the SMP effect. At $T = 0.5T_c$ (Fig. 6(e)), $J_c$ attains the maximum value at $x = 0.057$-$0.06$. Compared with $T = 5$ K, the light-blue area moves to lower $H$ region. For $T = 0.8T_c$ (Fig. 6(f)), the light-blue area further moves down to lower

$H$ region. Compared with K-Ba122, one can find a similarity that high $J_c$ values are attained in the under-doped side of the phase diagram and $J_c$ rapidly decreases with over-doping. On the other hand, distinct from K-Ba122, the SMP exists over the entire $x$-region, which fosters $J_c$ at finite $H$.

For $0.05 \leq x \leq 0.06$, $T_c$ and $H_{c2}$ increase with $x$. In this sense, the doping evolution of $J_c$ in the under-doped region can be attributed to the increase in the condensation energy, which leads to larger pinning force. On the other hand, $J_c$ shows an abrupt decrease between $x = 0.057$ and $x = 0.08$, even though $T_c$ and $H_{c2}$ are almost the same. This suggests that the character of the pinning centers abruptly changes between $x = 0.06$ and $0.08$.

### 3. P-Ba122

#### 1. Sample characterization

Fig. 7(a) shows the normalized $\chi(T)$ for the P-Ba122 single crystals with $x = 0.24, 0.30, 0.33, 0.38, 0.45$, and $0.52$. A sharp superconducting transition with $\Delta T_c \sim 0.5$-$1.0$ K was observed for the optimally- and the over-doped samples ($x = 0.30$-$0.45$). The under- ($x = 0.24$) and the heavily over-doped ($x = 0.52$) samples show broader transitions ($\Delta T_c \sim 2$-$3$ K). Fig. 7(i) shows the $x$ dependence of $T_c$. $T_c$ increases with $x$ up to $x = 0.30$ ($T_c^{\text{max}} = 29$ K) and decreases with further doping down to 15 K at $x = 0.52$. The $x$-dependence of $T_c$ is $dT_c/dx \sim 2$ K per percent P on the under-doped side and $\sim -0.8$ K on the over-doped side. These values are 50 % larger than for K-Ba122, but three to six times smaller than for Co-Ba122.

Fig. 7(b) shows $\rho_{ab}(T)$ (upper panel) and $d\rho_{ab}/dT$ (lower panel) below 150 K. $\rho_{ab}(T)$ of $x = 0.24$ shows an upturn at $T \sim 60$ K. A dip in $d\rho_{ab}/dT$ exists at $T \sim 50$ K as indicated by a black arrow, which corresponds to the PT-AFO phase transition. $T_{s/N}$ is also plotted in Fig. 7(i). For $x = 0.30$, such anomaly is absent while $\rho_{ab}(T)$ shows a slight deviation from the $T$-linear dependence below $T \sim 50$ K. For $x = 0.33$, $\rho_{ab}(T)$ shows a $T$-linear dependence for $T_c \leq T \leq 150$ K, which is often associated with the non-Fermi liquid charge transport, or quantum critical behavior [61,62].

Figs. 7(c)-(h) show $\rho_{ab}(T)$ for $H \mathbin{/\mkern-3mu/} c$ (top panels) and $H \mathbin{/\mkern-3mu/} ab$ (middle panels). The resistive transition is sharp ($\Delta T_c \sim 1$-$2$ K) except for $x = 0.24$ where $\Delta T_c \sim 8$ K. The transition shifts towards low $T$ with increasing $H$ without an appreciable resistive broadening. In the bottom panels, the $T$ dependences of $H_{c2}^c$ (filled circles) and $H_{c2}^{ab}$ (open circles) are plotted. $H_{c2}(0)$ for $H \mathbin{/\mkern-3mu/} c$ and $H \mathbin{/\mkern-3mu/} ab$ are estimated using the WHH formula and plotted against $x$ in Fig. 7(j). Note that $H_{c2}$ for $x = 0.24$ may not reflect the intrinsic values owing to the very broad transitions and to the apparent higher $T_c$ compared to that determined by $\chi(T)$. The values are marked in gray colors. $H_{c2}$ takes a maximum value of 50 T for $H \mathbin{/\mkern-3mu/} c$ and 110 T for $H \mathbin{/\mkern-3mu/} ab$ at $x = 0.30$, and decreases along the under- ($x = 0.24$) and over- ($x = 0.52$) doped sides. The maximum $H_{c2}$ is smaller than for K-Ba122 (170 T for $H \mathbin{/\mkern-3mu/} c$ and 300 T for $H \mathbin{/\mkern-3mu/} ab$) or Co-Ba122 (60 T for $H \mathbin{/\mkern-3mu/} c$ and 150 T for $H \mathbin{/\mkern-3mu/} ab$), while the maximum $T_c$ of 29.5 K is intermediate between K-Ba122 ($T_c^{\text{max}} = 38$ K) and Co-Ba122 ($T_c^{\text{max}} = 24$ K). As shown in Fig. 7(k), $\gamma$ takes values in the range of 1.6-2.5, again similar to K- and Co-Ba122.

#### 2. Magnetization hysteresis loops

Figs. 8(a)-(j) show the MHLs for P-Ba122 with $x = 0.24$-$0.45$ at selected temperatures (Figs. 8(a)-(e): 5 K $\leq T \leq 0.7T_c$, Figs. 8(f)-(j): $T \geq 0.7T_c$). Here, depending on the overall features, we classify them into two groups, namely, Group (1): under- and optimally-doped samples ($x = 0.24$ and $0.30$), where the magnitude of $M$ shows monotonous decrease with $T$ at any $H$, in other words, the MHLs at different $T$ do not cross each other, and Group (2): over-doped samples ($x = 0.33$ to $0.45$), where the MHLs at different $T$ cross each other. In the following the characteristics of the MHLs in each $x$ are compared.

Group (1) : For $x = 0.24$ (Figs. 8(a) and (f)), the MHLs are symmetric with respect to $M$ and $H$. At 5 K, a SMP is observed at $\mu_0 H_{\text{sp}} \sim 4$ T, which moves to lower $H$ with increasing $T$ and merges to the central peak at 9 K. For $x = 0.30$ (Figs. 8(b) and (g)), a clear SMP effect is observed, which persists up to $T = 28$ K $\sim 0.95T_c$. Compared with $x = 0.24$, the SMP is located at higher $H$. These behaviors are similar to under-doped K-Ba122 and Co-Ba122.

Group (2) : For $x = 0.33$ (Figs. 8(c) and (h)), the MHLs at different $T$ cross each other. With further doping, for $x = 0.38$ and $x = 0.45$, the shape of MHL becomes asymmetric with respect to the $M$ axis. Indeed for the $x = 0.45$ sample, at 13 K (black curve in Fig. 8(j)), $M$ takes the minimum value at $\mu_0 H \sim 2$ T and shows a step-like increase at $\mu_0 H \sim 2.5$ T on the $H$-increasing branch, while $M$ gradually decreases on the $\mu_0 H$-decreasing branch and takes its minimum at $\mu_0 H \sim 1.3$ T.

The above features are also seen in neutron-irradiated low-$T_c$ superconductors such as $V_3Si$ [63] and Nb [64], as well as in $MgB_2$ [65,66]. In these materials, weak pinning centers are introduced by the irradiation, and the SMP is associated with the order-disorder transition of the vortex lattice. When the neutron fluence is low, MHLs look similar to those of Group (2). With increasing neutron fluence, MHLs get similar to those of Group (1). Based on the similar features of the MHLs, the SMP in P-Ba122 can also be associated with the order-disorder transition. For P-Ba122, based on the MHL measurements, Salem-Sugui *et al.* already proposed that the vortex lattice structural phase transition is the origin of SMP [67]. If the order-disorder transition is indeed the case, the underlying mechanism of SMP is different from the vortex lattice structural phase transition. Similar features were also seen in K-Ba122 for $x = 0.33$ (Fig. 2(j)), also classified as Group (2).

So far, some groups reported the existence of the SMP effect [67,68] whereas other groups reported the its absence

[28,47]. In the latter cases, the MHLs were measured only in fields up to $\mu_0 H = 2$ T. According to the present results it is difficult to observe the SMP effect below 2 T, which may be the reason for the apparent discrepancy.

*3. Critical current density and vortex phase diagram*

Figs. 8(k)-(o) show the calculated $J_c$ for P-Ba122. The vertical scales differ from each other and red dashed lines correspond to $J_c = 0.1$ MA/cm$^2$. $J_c$ strongly depends on $x$. For the under-doped crystal, $x = 0.24$ (Fig. 8(k)), $J_c = 0.2$ MA/cm$^2$ was observed at $T = 5$ K and $\mu_0 H = 0$ T. $J_c$ rapidly falls off with increasing $H$, below 0.1 MA/cm$^2$ even at $T = 5$ K and $\mu_0 H = 0.5$ T. The largest $J_c = 1.0$ MA/cm$^2$ ($T = 5$ K and $\mu_0 H = 0$ T) was achieved for $x = 0.30$ (Fig. 8(l)). Among P-Ba122, only this composition sustains high $J_c$ values under magnetic fields up to high $T$. $J_c$ abruptly decreases from $x = 0.30$ (Group (1)) to 0.33 (Group (2)), while $T_c$ decreases only by 1 K. As shown in Fig. 8(m), $J_c = 0.5$ MA/cm$^2$ at $T = 5$ K and $\mu_0 H = 0$ T is about half of that for $x = 0.30$. Furthermore, at $T = 5$ K and $\mu_0 H = 6$ T, $J_c = 0.06$ MA/cm$^2$ is one fifth of that for $x = 0.30$. With further doping, for $x = 0.38$ and 0.45 (Figs. 8(n) and (o)), $J_c$ becomes further smaller.

Figs. 8(p)-(t) show the $T$ and $H$ dependence of $J_c$ by means of contour plots. The characteristic fields $H_{on}$ (pink circle), $H_{sp}$ (red diamond), $H_{irr}$ (yellow square), and $H_{c2}$ (orange triangle) are also plotted. Because all MHLs exhibit the SMP effect, the $H - T$ contour maps are separated into Region (I) to (IV) for all $x$s.

For $x = 0.24$ (Fig. 8(p)), a relatively high-$J_c$ (light-blue) area covers only the low $T$ region. Here, $H_{c2}(T)$ is omitted owing to the broad resistive transition. For $x = 0.30$ (Fig. 8(q)), a high-$J_c$ area extends over a wide $T$ and $H$ range. $H_{sp}$ is larger compared with $x = 0.24$, which results in the increase of Region (II), also producing a high-$J_c$ area around it up to high $H$. Region (IV) covers a narrow area, reflecting that $H_{irr}$ and $H_{c2}$ are close to each other.

For $x = 0.33$ shown in Fig. 8(r), $H_{sp}$ line lies closer to $H_{irr}$, possibly reflecting the decrease in the pinning energy. Region (II) does not support high $J_c$. With further doping, the phase diagrams of $x = 0.38$ (Fig. 8(s)) and 0.45 (Fig. 8(t)) are dominated by the low-$J_c$ (blue-color) area except for zero field. $H_{sp}$ locates close to $H_{irr}$, which, according to the order-disorder transition model, suggests a further decrease in the pinning energy with over-doping.

*4. Doping dependence of critical current density*

Figs. 9 (a)-(c) show $J_c(H)$ of P-Ba122 with various $x$s at $T = 5$ K, $0.5T_c$, and $0.8T_c$, respectively. The $x$-dependence of $J_c$ is visualized in Figs. 9 (d)-(f) by means of a contour plot. At 5 K (Fig. 9(d)), starting from $x = 0.24$, $J_c$ increases with $x$ and attains the maximum value at $x = 0.30$. Around $x = 0.30$, high $J_c$ is sustained up to 7 T. $J_c$ rapidly decreases toward $x = 0.33$. At $T = 0.5T_c$ (Fig. 9(e)), the light-blue area emerges at high $\mu_0 H \sim 6$ T in the over-doped region, reflecting the SMP located close to $H_{irr}$. For $T = 0.8T_c$ (Fig. 9(f)), the light-blue area moves down to the lower $H$ region.

## 4. Doping dependence of $J_c$ for three cases

Figs. 10(a)-(f) summarize the $x$-dependence of $J_c$ for K-, Co-, and P-Ba122 at $\mu_0 H = 0.4$ T ((a)-(c)) and 5 T ((d)-(f)) in form of contour plots. In the same Figures, the $T_c$s are also plotted using open circles. Here, the red-color regions correspond to $J_c \geq 1$ MA/cm$^2$ in Figs. 10(a)-(c) (0.4 T) and $J_c \geq 0.7$ MA/cm$^2$ in Figs. 10(d)-(f) (5 T), respectively. In general, K-Ba122 possesses higher $J_c$ compared with Co- and P-Ba122. At $\mu_0 H = 0.4$ T, the highest $J_c$ of 2.6 MA/cm$^2$ at 5 K is achieved in K-Ba122 at $x = 0.30$. Co-Ba122 possesses the highest $J_c$ of 0.9 MA/cm$^2$ at $x = 0.057$, being a little bit above the highest $J_c$ of P-Ba122, 0.7 MA/cm$^2$ at $x = 0.30$. The same tendency is observed at higher field $\mu_0 H = 5$ T; the highest $J_c$ values at 5 K are 0.7 MA/cm$^2$ in K-Ba122 ($x = 0.30$), 0.5 MA/cm$^2$ in Co-Ba122 ($x = 0.057$), and 0.2 MA/cm$^2$ in P-Ba122 ($x = 0.30$), respectively. Judging from $J_c$ of the pristine single crystal samples, K-Ba122 is superior to P- and Co-Ba122 as a candidate material for applications.

Common to the three doping variations, $J_c$ increases with $x$ from the under- to the optimally-doped region, apparently following the increase in $T_c$. With further doping, $J_c$ rapidly decreases above critical doping levels, $x_c \sim 0.30$ for K-Ba122, $\sim 0.06$ for Co-Ba122, and $\sim 0.30$ for P-Ba122, respectively. Figs. 10(a) and (d) also confirm that for K-Ba122, the $x$-dependence of $J_c$ is distinct from that of $T_c$. For Co- and P-Ba122, highest $J_c$ is realized nearly at the highest $T_c$ compositions, $x \sim 0.06$ for Co-Ba122 and $x \sim 0.30$ for P-Ba122, respectively. However, on the over-doped side of the phase diagram, $J_c$ decreases more rapidly in contrast to the weak decrease in $T_c$, as indicated by the large blue areas spreading in the over-doping regions. In this regard, the doping dependence of $J_c$ does not simply follow $T_c$ in all three cases.

## IV. DISCUSSION

## 1. Magnetic-field dependence of pinning force density

In this section, in order to gain insight into the pinning mechanism, we analyze the pinning force density, defined as $F_p = J_c \times \mu_0 H$. Fig. 11 shows the pinning force density normalized to its maximum value, namely, $f_p = F_p/F_p^{max}$ is plotted against the reduced magnetic field, $h = H/H_{irr}$, for K-Ba122 (Figs. 11 (a) - (d)), for Co-Ba122 (Figs. 11(f)-(i)), and for P-Ba122 (Figs. 11(k)-(n)), respectively. For most samples, $H_{irr}$ exceeds the accessible magnetic field of 7 T at low $T$. In such cases only data referring to high $T$ are presented. In general, $f_p(h)$ follows the functional form, $f_p(h) = Ah^p(1-h)^q$, where $A$ is a constant, and $p$ and $q$ are parameters providing information of the pinning mechanism

[69]. If pinning is governed by a single mechanism within a certain $T$-range, the $f_p(h)$ at different $T$ collapse into one master curve, in other words, scaling of $f_p(h)$ as a function of $h$ is expected. It has to be mentioned that the Dew-Hughes model [69] is based on purely geometric arguments neglecting the elasticity of the vortex lattice. This means that it is valid in a plastically deformed lattice only and in principle inappropriate for discussing the pinning mechanisms in samples showing the SMP, where the elastic and pinning energies are comparable and effectively compete with each other. In particular, the position of the SMP resulting from an order-disorder transition of the vortex lattice is a function of the defect density (or pinning force) and even not necessarily constant in temperature [58]. A shift of the maximum in $J_c$ obviously results in a shift of the maximum in the pinning force; thus the peak of $f_p$ can be at different positions irrespective of the pinning mechanism. However, this kind of analysis is frequently done in literature and we will apply it to our data for the sake of comparison with literature and between the different samples. While a universal form of $f_p$ at various temperatures is certainly a strong indication for one underlying pinning mechanism, a shift of the peak is not necessarily caused by a change of the pinning mechanism.

Figs. 11(a)-(d) show $f_p(h)$ for K-Ba122 with $x = 0.23$, 0.30, 0.33, and 0.41, respectively. For $x = 0.23$ (Fig. 11(a)), $f_p(h)$ between 16 K and 20 K lie on a single curve, indicating that a single pinning mechanism plays a dominant role within this $T$-range. $f_p(h)$ exhibits a peak at $h_{max} \sim 0.39$. For $x = 0.30$ (Fig. 11(b)), which possesses the highest-$J_c$, the peak position $h_{max}$ shifts to $\sim 0.46$, suggesting a change in the dominant pinning mechanism, or a reduction of the defect density, $\rho_D$. $h_{max}$ becomes larger, up to 0.56 for $x = 0.33$ (Fig. 11(c)), likely due to a further reduction of $\rho_D$, then suddenly decreases down to 0.20 for $x = 0.41$ (Fig. 11(d)). The latter change is associated with the disappearance of the SMP effect, pinning is too weak to disorder the flux line lattice at high fields. The $x$-dependence of $h_{max}$ is shown in Fig. 11(e), together with the $x$-dependence of $F_p^{max}$ at $T = 0.9T_c$. Apparently, the $x$-dependence of $h_{max}$ correlates with that of $F_p^{max}(0.9T_c)$ and highlights the sudden change in pinning which occurs between $x = 0.33$ and 0.36. The present results are consistent with previous reports [24,41].

Figs. 11(f)-(i) show $f_p(h)$ for Co-Ba122 with $x = 0.05$, 0.06, 0.08, and 0.10. For $x = 0.05$ (Fig. 11(f)), $f_p(h)$ shows a reasonable scaling behavior with $h_{max} \sim 0.32$ at low $T$ between 7 K and 9 K. However, at 10 K and 11 K, $f_p(h)$ deviates from the scaling behavior, suggesting that different pinning mechanisms are at work. In the case of $x = 0.06$ (Fig. 11(g)), a scaling behavior is observed above 17 K ($\sim 0.7T_c$). Here $h_{max}$ is 0.38, which is similar to, but slightly larger than that of $x = 0.05$. A similar scaling behavior is observed for the over-doped samples, $x = 0.08$ (Fig. 11(h)) and $x = 0.10$ (Fig. 11(i)), with $h_{max} = 0.42$-0.44. The $x$-dependence of $h_{max}$ and $F_p^{max}(0.9T_c)$ are plotted in Fig. 11(j). Besides the heavily under-doped region ($x \sim 0.05$), $h_{max}$ shows weak $x$-dependence, suggesting that the relevant pinning landscape does not change with $x$. Presumably related to the constant $h_{max}$, the $x$-dependence of $F_p^{max}(0.9T_c)$ is rather gradual, in contrast to K-Ba122 in which both $F_p^{max}$ and $h_{max}$ significantly change with $x$.

$f_p(h)$ for P-Ba122 with $x = 0.24$, 0.30, 0.33, and 0.45 are shown in Figs. 11(k)-(n). For $x = 0.24$ (Fig. 11(k)), $h_{max}$ changes with $T$, from $h_{max} = 0.45$ at 9 K to $h_{max} = 0.2$ at 13 K. The behavior resembles $x = 0.05$ Co-122 (Fig. 11(f)) and indicates that the dominant pinning source changes with $T$. For $x = 0.30$ (Fig. 11(l)), the scaling behavior prevails between 17 K and 22 K with $h_{max}$ of 0.45. The same $h_{max}$ value is obtained also for $x = 0.24$ at low $T$, presumably suggesting the same pinning sources. For $x = 0.33$ (Fig. 11(m)), the scaling is demonstrated between 23 K and 27 K. Here $h_{max}$ increases up to 0.62, indicative of a weakening in the pinning, consistent with the decrease of $F_p^{max}$. $h_{max}$ further increases up to 0.7 for $x = 0.45$ (Fig. 11(n)), and the peak width becomes narrower, both consistent with the order-disorder scenario and a pinning structure which can disorder the vortex lattice only close to $H_{irr}$. Furthermore, at $T > 16$ K, an extra feature appears at low $h$ region. The present observations, namely, multiple features in $f_p(h)$ and large $h_{max} \sim 0.7$ are in good agreement with the report by Fang et al. [68] The $x$-dependences of $h_{max}$ and $F_p^{max}(0.9T_c)$ are plotted in Fig. 11(o). The $x$-dependence of $h_{max}$ in P-Ba122 resembles the neutron-fluence dependence of $h_{max}$ in neutron-irradiated superconductors, which can be understood in terms of the order-disorder transition with different density of pinning centers.

Through the comparison of $f_p(h)$ for K-, Co-, and P-Ba122, several similarities and differences can be pointed out. (1) For under-doped Co-Ba122 (Fig. 11(a)) and P-Ba122 (Fig. 11 (k)), scaling behavior is violated at high $T$. This may indicate the existence of multiple pinning sources which become dominant at different $T$. (2) For the optimally-doped samples, scaling behavior prevails at least at high $T$. $h_{max}$ tends to increase with increasing $x$ and the highest $J_c$ is attained when $h_{max}$ becomes 0.40-0.45, at $x = 0.30$ for K-Ba122 (Fig. 11 (b)), $x = 0.06$ for Co-Ba122 (Fig. 11(g)), and $x = 0.30$ for P-Ba122 (Fig. 11(l)), respectively. The coincidence suggests that highest $J_c$ is caused by the same pinning mechanism(s) common to the three samples. (3) In the over-doped region, $f_p(h)$ behaves differently (Figs. 11(d), (i), and (n)). This suggests that the dominant pinning mechanism which produces high $J_c$ disappears in the over-doped region and other pinning mechanisms which are more specific to the dopant variation become at work.

The scaling analysis presented above suggests that a single pinning mechanism is dominant over a certain $T$-range. As stated, however, one should keep in mind that the scaling analysis can be accurately performed only if $H_{irr}$ is well-defined. At low $T$, $H_{irr}$ exceeds the upper limit of the accessible magnetic field of 7 T, and therefore it is not clear at this moment whether there exist extra pinning mechanisms at lower $T$. Violation of the scaling behavior in the under-and over-doped Co-Ba122 and P-Ba122 indeed suggests the existence of multiple pinning sources. In the next section, we investigate the $T$-dependence of $J_c$ in order to complement the $x$- and $T$-dependence of the pinning mechanism.

## 2. Temperature dependence of $J_c$

In discussing the $T$-dependence of $J_c$, one should consider two distinct contributions, i.e. strong pinning and weak collective pinning. The strong pinning comes from sparse and large (nm-sized) heterogeneities, while the weak collective pinning is related to dense, small (atomic-scale) pinning centers. Several results indicate the existence of strong pinning. For example, small-angle neutron scattering [70], Bitter decoration [71], magnetic force microscopy [70], and scanning tunneling microscopy [72] studies revealed a disordered vortex lattice. Here the vortex lattice is expected to be highly disordered because each vortex is preferentially pinned by the sparse, randomly distributed pinning sites. Also, the magnetization hysteresis loop measurements show the sharp peak at around zero field and the power-law decay of $J_c$ ($J_c \sim H^{-\alpha}$ ($\alpha \sim 0.5$)) at low-field region, which are attributed to the strong pinning contribution [27,28].

On the other hand, it has been also shown that the overall pinning properties of Fe-based superconductors have been successfully explained in terms of collective pinning and vortex creep [29-34,45,73]. Based on these considerations, in this section, we apply the collective pinning and creep model to our results and extract the trend in the doping-dependent $J_c$. It is pointed out that the strong pinning contribution rapidly decreases in $H$ ($J_c \sim H^{-\alpha}$) and the weak collective pinning contribution becomes dominant with increasing $H$ ($> 0.1$-$1$ T), therefore, one can approximate the $T$-dependence of $J_c$ at high $H$ by collective pinning contributions.

At around zero field and at higher fields, $J_c$ is dominated by the self-field effect [74,75] and the SMP effect, respectively, which hinders a meaningful application of collective pinning theory in these regions. On the other hand, collective pinning theory is applicable in the moderate $H$ regions, when thermally activated flux motion and collective pinning / creep [76] are to be taken into account. Two types of pinning mechanisms, namely, $\delta T_c$ pinning caused by the spatial variation of $T_c$, and $\delta l$ pinning caused by the fluctuation in the mean free path ($l$) [42], have to be considered. The corresponding $T$-dependence of $J_c$ is then described by the following formula [44,77], using the reduced temperature $t = T/T_c$,

$$J_c(t) = J_{c0}J(t)/\{1 + [\mu k_B TC/U_0 g(t)]\}^{1/\mu},$$

where $J_{c0}$ is $J_c$ at $T = 0$ K, $\mu$ is the glassy exponent characterizing the vortex creep regime (single vortex, small bundle, etc.), $k_B$ is the Boltzmann constant, $U_0$ is the pinning potential, and $C$ is a constant related to the hopping attempt frequency. For $H < H_{on}$, it is known that $\mu$ can be treated as a constant [30,78]. Assuming the single-vortex pinning, the $T$-dependent terms, $J(t)$ and $g(t)$, are expressed as [79]

$$J(t) = (1 - t^2)^{7/6}(1 + t^2)^{5/6},$$
$$g(t) = (1 - t^2)^{1/3}(1 + t^2)^{5/3}$$

for $\delta T_c$ pinning, and

$$J(t) = (1 - t^2)^{5/2}(1 + t^2)^{-1/2},$$
$$g(t) = 1 - t^4$$

for $\delta l$ pinning, respectively. Based on the collective pinning model [42], the condition that the above formulas are applicable is $H < B_{SB} \sim 5(J_c/J_d)H_{c2}$, where $J_d$ is the depairing current. In the case of K-Ba122 ($x = 0.30$), $B_{SB}$ is estimated as $\sim 1.5$ T using $J_c \sim 6 \times 10^5$ A/cm$^2$, $J_d \sim 2 \times 10^8$ A/cm$^2$ [80], and $\mu_0H_{c2} \sim 100$ T. Similarly, $B_{SB}$ is $\sim 1.6$ T for Co-Ba122 ($x = 0.06$) and $\sim 1.4$ T for P-Ba122 ($x = 0.30$), respectively. For the present analysis, $J_c(T)$ at $\mu_0H = 0.4$ T are employed.

$J_c(t)$ caused by $\delta T_c$ pinning ($J_c^{\delta T_c}(T)$) and $J_c(t)$ caused by $\delta l$ pinning ($J_c^{\delta l}(T)$) exhibit different $T$-dependences. Particularly, at intermediate $T$, $J_c^{\delta l}(T)$ decreases much faster than $J_c^{\delta T_c}(T)$. In order to extract the contributions of the two pinning mechanisms, we fitted the experimental results using a function $J_c(T) = J_c^{\delta T_c}(T) + J_c^{\delta l}(T)$. It should be noted that the formula is not exact since the sum of the two contributions are not necessarily additive. Nevertheless, as shown below, this simple formula successfully picks up the dominant pinning mechanism, particularly when one overwhelms the other, as well as the doping evolution of the two contributions. To fit data, we used four parameters; $\mu$, $k_BC/U_0$ (K$^{-1}$), $J_{c0}^{\delta T_c}$ ($= J_c^{\delta T_c}(0)$) (MA/cm$^2$), and $J_{c0}^{\delta l}$ ($= J_c^{\delta l}(0)$) (MA/cm$^2$). For $\mu$ and $k_BC/U_0$, we took the reported values obtained from the magnetization-relaxation measurements [30,78]. Note that we employed the same $\mu$ and $k_BC/U_0$ for $J_c^{\delta T_c}(T)$ and $J_c^{\delta l}(T)$, which are not necessarily the same for different pinning mechanisms. This is because, when one mechanism is dominant, $J_c(T)$ is mostly described by either $J_c^{\delta T_c}(T)$ or $J_c^{\delta l}(T)$, therefore the values of $\mu$ and $C/U_0$ represent the dominant mechanism and the minor contribution does not affect the results. Indeed, as shown in the APPENDIX, the main results do not depend on the magnitude of these parameters.

Figs. 12(a)-(d) show $J_c(T)$ for K-Ba122 plotted in a semi-logarithmic scale. Here black circles are experimental results. Dashed curves are the results obtained through the fitting. The blue-dashed and the red-dashed curves are the contributions from the $\delta T_c$ pinning and the $\delta l$ pinning, and the black-dashed curves are their sums, respectively. For $x = 0.23$, the $\delta l$ pinning is larger than the $\delta T_c$ pinning at low $T$. With increasing $T$, the $\delta T_c$ pinning increases, and becomes dominant at $T > 10$ K.

In the case of $x = 0.30$ (Fig. 11(b)) which possesses the highest $J_c$ among K-Ba122, $J_c(T)$ decreases linearly over a wide $T$-range between 5 K and 20 K, followed by a steeper fall-off at higher $T$. This is the typical behavior of $\delta l$ pinning. Indeed, $J_c(T)$ is fitted well by $\delta l$ pinning alone up to 20 K, with a significant contribution from $\delta T_c$ pinning only close to $T_c$. The fitting parameters $\mu$, $k_BC/U_0$, and $J_{c0}^{\delta T_c}$ are the same as those used for $x = 0.23$, whereas $J_{c0}^{\delta l}$ is larger by one order of magnitude.

Fig. 12(c) shows $J_c(T)$ for $x = 0.33$. The overall $T$-dependence is similar to that for $x = 0.22$ rather than for $x = 0.30$, in the sense that it shows concave $T$-dependence, with a hump at $T \sim 25$ K. The hump feature is characteristic for $\delta T_c$ pinning. The fitting result shows that the $\delta l$ pinning contributes the most below 15 K, while $\delta T_c$ pinning becomes dominant at higher $T$. In this case, a small $\mu$ value reproduces the experimental results, which indicates a faster flux creep.

For $x = 0.51$, $J_c(T)$ is reproduced well by $\delta T_c$ pinning alone and the contribution from the $\delta l$ pinning is smaller by more than one order of magnitude even at the lowest $T$.

The obtained results indicate that the dominant pinning mechanism of K-Ba122 changes depending on $T$ and $x$. To see the $x$-dependences of the two pinning contributions, $J_c^{\delta T_c}$ and $J_c^{\delta l}$ at $T = 0.3T_c$ are plotted in Fig. 13(a). The contribution from $\delta T_c$ pinning (blue circles) shows a modest $x$-dependence, with a maximum at $x = 0.25$. On the other hand, the contribution from the $\delta l$ pinning exhibits a pronounced peak at $x = 0.30$. The present analyses suggests that the strong enhancement in $J_c$ for $x = 0.30$ is mainly due to strong $\delta l$ pinning which is special to this composition.

Figs. 12(e)-(h) show $J_c(T)$ for Co-Ba122. For $x = 0.05$, shown in Fig. 12(e), the $\delta T_c$ pinning is dominant for the entire $T$-range, while $\delta l$ pinning makes certain contributions at low $T$, up to one third of the total $J_c$. Fig. 12(f) shows $J_c(T)$ for $x = 0.057$, which possesses the highest $J_c$ among Co-Ba122. Here the experimental data exhibit convex $T$-dependence. This behavior resembles $x = 0.30$ K-Ba122 and indicates that $\delta l$ pinning plays a dominant role, although the contribution from $\delta T_c$ pinning is pronounced near $T_c$. For the over-doped sample with $x = 0.08$ shown in Fig. 12(g), $J_c(T)$ shows a concave $T$-dependence at $\sim 8$ K, which is characteristic for $\delta T_c$ pinning. The result indicates that the $\delta l$ pinning significantly weakens between $x = 0.057$ and $x = 0.08$. For $x = 0.10$ (Fig. 12(h)), $J_c(T)$ is reasonably fitted by the $\delta T_c$ pinning contribution alone. In Fig. 13(b), the $x$-dependence of $J_c^{\delta T_c}$ and $J_c^{\delta l}$ of Co-Ba122 at $T = 0.3T_c$ is plotted. As in the case of K-Ba122, the $\delta T_c$ pinning is enhanced in the under-doped region and the $\delta l$ pinning shows a sharp peak at $x \sim 0.06$. It is also noted that the contribution from $\delta T_c$ pinning is larger for Co-Ba122 compared to K-Ba122.

$J_c(T)$ for P-Ba122 is shown in Fig. 12(e)-(h). For $x = 0.24$ (Fig. 12(i)), the experimental data are well reproduced by $\delta T_c$ pinning alone. On the other hand, for $x = 0.30$ which possesses the highest $J_c$ among P-Ba122, $J_c(T)$ shows a linear $T$-dependence between 5 K and 15 K, and the slope becomes steeper at higher temperatures, which is the characteristic of $\delta l$ pinning, similar to the high-$J_c$ K- and Co-Ba122 samples (Figs. 12(b) and (f)). The fitting result indeed shows that $\delta l$ pinning is dominant at low $T$, while $\delta T_c$ pinning is pronounced near $T_c$. The $J_c(T)$ curve of the slightly over-doped sample with $x = 0.33$ (Fig. 12(k)) can be expressed as a sum of the $\delta T_c$ pinning and the $\delta l$ pinning contributions with a ratio similar to that for $x = 0.30$. The result indicates that both contributions are equally suppressed from $x = 0.30$ to $x = 0.33$. For $x = 0.45$ (Fig. 12(l)), the magnitude of the two contributions decrease by one order of magnitude compared with $x = 0.33$. Here the $\delta T_c$ pinning contribution is larger for the whole temperature range. Fig. 13(c) depicts the $x$-dependence of $J_c^{\delta T_c}$ and $J_c^{\delta l}$ of P-Ba122 at $T = 0.3T_c$. In this case, both $\delta T_c$ pinning and $\delta l$ pinning components show a peak at $x = 0.30$.

To summarize, common to the three cases, a significant enhancement of the $\delta l$ pinning is observed for high-$J_c$ samples, particularly at low $T$. Also, the $\delta T_c$ pinning, which becomes important at high $T$, tends to be enhanced on the under-doped side. Both contributions significantly decrease on the over-doped side.

Note that the present fitting based on the collective pinning model may not be adequate for several samples. Also, $J_c(T)$ data at high $T$ tend to be affected by the SMP which shifts to lower $H$ with increasing $T$, leading to the overestimation of the $\delta T_c$ pinning. However, this does not affect our conclusion because (1) $J_c$ of those cases is one order of magnitude smaller than for the high-$J_c$ samples and (2) we found a significant enhancement of $\delta l$ pinning to be responsible for high $J_c$.

## 3. Possible sources for $\delta T_c$ pinning and $\delta l$ pinning

In this section, we discuss the possible sources for $\delta T_c$ and $\delta l$ pinning. By definition, $\delta T_c$ pinning is caused by the spatial variation in $T_c$. In the Ba122 system, $T_c$ is determined by the concentration of dopant atoms, i.e. $x$. In such a situation, the spatial variation in $T_c$ ($\Delta T_c$) is introduced by the spatial inhomogeneity of the dopant atom distribution and is expressed as

$$\Delta T_c = |dT_c/dx| \Delta x,$$

where $dT_c/dx$ is the slope of the $T_c(x)$ curve, and $\Delta x$ is its spatial variation in $x$. The formula implies that $\Delta T_c$ becomes large if $T_c$ strongly depends on $x$ ($dT_c/dx$ is large). On the other hand, the $\delta l$ pinning is associated with the spatial variation of the mean free path ($l$) of the charge carriers. Experimentally, $l$ is related to the resistivity $\rho(T)$ by the following formula

$$\rho(T) = (\hbar/e^2)(3\pi^2/n_0^2)^{1/3}/l,$$

where $n_0$ is the carrier density. In particular, the residual resistivity ($\rho_0$), the $T$-independent part of the normal state resistivity, directly reflects $l$ associated with the spatial inhomogeneity.

The $x$-dependence of $J_c^{\delta T_c}$ and $J_c^{\delta l}$ for K-, Co-, and P-Ba122 are shown in Figs. 13(a)-(c). In Figs. 13(d)-(f), we show the $x$-dependence of $|dT_c/dx|$ (blue dashed line) derived from the $T_c(x)$ curve (black dashed line). $|dT_c/dx|$ becomes zero at $T_c^{max}$; $x = 0.36$ for K-Ba122, $x = 0.06$ for Co-Ba122, and $x = 0.31$ for P-Ba122, respectively. In all cases, the $T_c(x)$ curve is more inclined on the under-doped side compared with the over-doped side. As a result, $|dT_c/dx|$ is always larger on the under-doped side. This tendency is consistent with enhanced $\delta T_c$ pinning in the under-doped region as seen in Figs. 13(a)-(c). Moreover, $|dT_c/dx|$ of Co-Ba122 is larger than for K- and P-Ba122 and persists up to the over-doped region, which likely enhances the $\delta T_c$ pinning in Co-Ba122. Assuming the local variation in $T_c$ caused by the inhomogeneous distribution of dopant elements, the doping and dopant dependence of the $\delta T_c$ pinning can be well explained.

Next, we discuss the possible relationship between $\rho_0$ and the $\delta l$ pinning. In Figs. 13(g)-(i), we show the $x$-dependence of $\rho_0$ in K-, Co-, and P-Ba122. Here $\rho_0$ is estimated by fitting $\rho_{ab}(T)$ using the formula $\rho_{ab}(T) = \rho_0 + AT^n$ at low $T$, typically below 80 K. Examples of the fitting results are shown in the insets. For all the cases, $\rho_0$ is large in the under-doped samples, and rapidly decreases with $x$ toward the optimally-doped region. In the over-doped region, $\rho_0$ does not depend on $x$. In the optimally- and the over-doped region, there is a correlation between $\rho_0$ and $J_c^{\delta l}$, namely, both quantities are small in the over-doped region, and towards the optimally-doped region, they sharply increase at the same $x$. In the heavily under-doped region, the two quantities exhibit different $x$-dependence. With decreasing $x$, $J_c^{\delta l}$ begins to decrease, whereas $\rho_0$ continues to increase. It may be because of the weakening of the pinning energy due to the decrease in condensation energy.

The reason for the enhanced $\rho_0$ in the under-doped region is not clear at this moment. Impurity scattering from the dopant atoms themselves may not be the origin, because $\rho_0$ decreases with increasing the number of dopant atoms. This tendency is opposite to what we expect from the impurity scattering. Furthermore, K atoms are located away from the FeAs planes and thus should not be effective scattering centers compared with Co atoms which directly substitute Fe atoms. This contradicts with the fact that $\rho_0$ of under-doped K-Ba122 is comparable with those for Co- and P-Ba122. One possible scenario is that in the heavily under-doped region, the AFO phase coexists with the SC phase in a microscopic length scale in the under-doped region as has been suggested by the muon spin rotation, nuclear magnetic resonance, *etc*. [81-83] In such a situation, the material become microscopically inhomogeneous, which results in limiting $l$. The $\delta T_c$ pinning is also expected to increase, since the AFO phase is a non-SC, and thus works as $\delta T_c$ disorder.

There are several proposals which highlight the AFO-PT phase transitions. Based on the noticeable enhancement in $J_c$ in slightly under-doped Co-Ba122, Prozorov *et al.* proposed that the twin domain boundaries of the orthorhombic phase act as pinning centers [40]. Later, Kalisky *et al.* carried out the scanning SQUID susceptometry on Co-Ba122 and reported that the superfluid density is enhanced on the twin boundaries [84]. The authors suggest that the enhancement in the superfluid density at the twin boundary results in an enhanced $J_c$ [85]. The phase diagram of the K-Ba122 (Fig. 9 (i)), Co-Ba122 (Fig. 5(i)), and P-Ba122 (Fig. 9(i)) obtained in the present studies are compatible with these proposals, in that highest $J_c$ compositions nearly coincide with the critical concentration for the AFO transition in all three cases. It is also likely that the twin domain boundary acts as a scattering center and causes the $\delta l$ pinning. Likewise, an inhomogeneous superfluid density produces the $\delta T_c$ pinning. Measurements on the detwinned single crystals would answer the question whether the twin domain boundary is indeed responsible for the enhanced $J_c$.

The above arguments are more or less based on the conventional viewpoints. More exotic scenario which is related to the quantum criticality has been proposed by Putzke *et al*. [86] Based on the detailed upper- and lower-critical field measurements on P-Ba122, they proposed that the energy of superconducting vortices is enhanced near the possible quantum critical point (QCP) at $x \sim 0.3$, possibly due to a microscopic mixing of antiferromagnetism and superconductivity. Based on the scenario, the vortex state of the Fe-based superconductors is highly unusual. Their results are apparently consistent with the spike-like enhancement in $J_c$ observed in the present study. On the other hand, if the scenario is correct, one would naturally expect that the QCP also exists in K-Ba122 and Co-Ba122. To our knowledge, there is no confirming experimental evidence which support the QCP in these materials.

Toward the applications, the present results suggest that the introduction of $\delta l$ pinning disorder is an effective way to enhance $J_c$. It may be possible that the introduction of the $\delta l$ pinning into the highest $T_c$ sample results in a material possessing both highest $T_c$ and highest $J_c$. In this regard, co-doping of Co into optimally-doped K-Ba122 would be intriguing.

Finally, we briefly compare the present results with another candidate for applications, i.e. 1111 system, which possesses the highest $T_c^{max} \sim 56$ K and a relatively large anisotropy $\gamma \sim 5$-$10$. Reflecting the larger $\gamma$, a significant broadening of resistive transition under $H$ is observed in the 1111 system [87]. The reported $J_c \sim 2$ MA/cm$^2$ at 5 K and 0 T in SmFeAsO$_{0.7}$F$_{0.25}$ ($T_c \sim 50$ K) [11] is comparable with that of K-Ba122 while $T_c$ is higher by ~12 K, suggestive of a weaker pinning. Indeed, it is reported that the vortex dynamics in 1111 system is governed by a Josephson-like vortex behavior at low $T$s, leading to a weak interaction between vortices and pinning sites [88]. These features indicate that nature of pinning in 1111 system is different from 122 system. At this moment, the doping dependence of $J_c$ in 1111 system is unclear. It is expected that the doping dependence of $J_c$ is different from 122 system owing to the larger anisotropy and the consequent difference in nature of pinning. On the other hand, because the electronic phase diagram of 1111 system is more or less similar to that of 122 system, a larger $J_c$ may be achieved through the optimization of doping level, presumably around slightly under-doped region, as demonstrated in this work.

## V. CONCLUSIONS

In this study, we studied the dependence of $J_c$ for K-, Co-, and P-Ba122 on the dopant concentration using high-quality single crystals and established the doping dependence of $J_c$. $J_c$ shows a variety of $H$- and $T$-dependences depending on the variation and concentration of dopant elements. On the other hand, in all the cases, the magnitude of $J_c$ is sharply enhanced at doping levels corresponding to the slightly under- to optimally-doped region. The common enhancement of $J_c$ in spite of the distinct character (charge type and the substitution site) of dopants indicates that the behavior comes from an intrinsic origin, which is likely related to the underlying electronic phase diagram. The analysis of $f_p(h)$ showed a similar $h_{max}$ value for high-$J_c$ samples, suggesting that a common pinning mechanism is responsible for enhancing $J_c$. Based on the $T$-dependence of $J_c$, it was found that both the $\delta T_c$ pinning and the $\delta l$ pinning are enhanced in the under-doped region. The extracted results are consistent with the larger $|dT_c/dx|$ as well as the increase of $\rho_0$ in the under-doped side, which indicate enhanced $\delta T_c$ and $\delta l$ disorder, respectively.

## ACKNOWLEDGEMENTS

This work was supported by the Austrian Science Fund (FWF): P22837-N20, the European-Japanese collaborative project SUPER-IRON (No. 283204) of the Japan Science and Technology Agency (JST), Austria-Japan Bilateral Joint Research Project hosted by Japan Society for the Promotion of Science (JSPS) and by the Austrian Science Fund (FWF): I2814-N36, and a Grant-in-Aid for Scientific Research (KAKENHI) (No. 16767447) from JSPS. D.-J. S. thanks JSPS for the financial support.

## APPENDIX

### 4. Fitting-parameter dependence of $\delta T_c$ and $\delta l$ pinning contributions

In Fig. 12 in Sec. IV B, we showed results of fitting our data by using the parameters shown in each panel. Because we did not obtain $\mu$ and $C/U_0$ experimentally, these parameters can be arbitrarily chosen although we referred to reported values. In fact, one can fit the data by using different parameter sets, possibly resulting in a considerable uncertainty arising from the choice of these parameters. In order to check how much the results depend on the choice of the parameters, we varied $\mu$ in the range of 0.2-1.2 (which covers the reported values for iron pnictides) and carried out the fitting. Figs. 14(a)-(l) show the $\mu$-dependence of the $\delta T_c$ and $\delta l$ pinning contributions (the derived $k_B C/U_0$ values are also plotted). The samples correspond to those used in Fig. 12. In most cases, the dominant contribution does not change for any $\mu$ values. For several samples (Figs. 14(a), (k), and (l)), where the contributions from $\delta T_c$ and $\delta l$ pinning are comparable, the dominant contribution changes when $\mu$ is varied, although they are still comparable with each other. The results show that the overall doping dependence of $\delta T_c$ and $\delta l$ pinning contributions is robust against the choice of $\mu$, hence it does not change our conclusion. The error bars in Figs. 13(a)-(c) indicate the uncertainty arising from the $\mu$-dependence.

# FIGURES

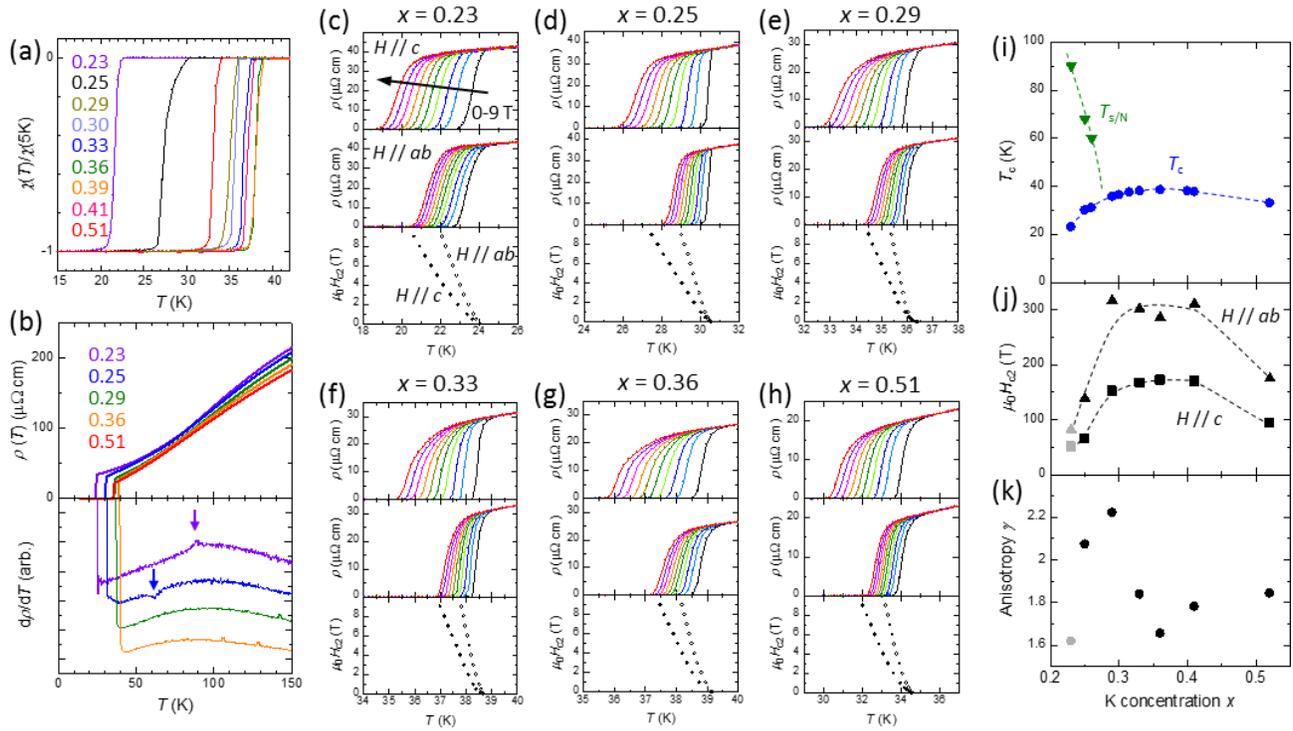

FIG. 1. (Color online) Characterization of Ba$_{1-x}$K$_x$Fe$_2$As$_2$ single crystals with $x$ = 0.23-0.51. (a) $T$ dependence of magnetization. (b) $T$ dependence of in-plane resistivity (upper panel) and its derivative (lower panel). (c)-(h) In-plane resistivity in magnetic fields parallel to $c$ axis (upper panel) and $ab$ plane (lower panel). The bottom panel show $H_{c2}$ with $H // c$ (filled circles) and $ab$ (open circles) determined by 90% of normal state resistivity. (i)-(k) Doping dependence of $T_c$ and $T_{s/N}$, $H_{c2}(0)$ for $H // c$ (filled circles) and $ab$ (open circles), and anisotropy $\gamma$ of $H_{c2}$ ((d$H_{c2}^{ab}$/d$T$)/(d$H_{c2}^{c}$/d$T$)).

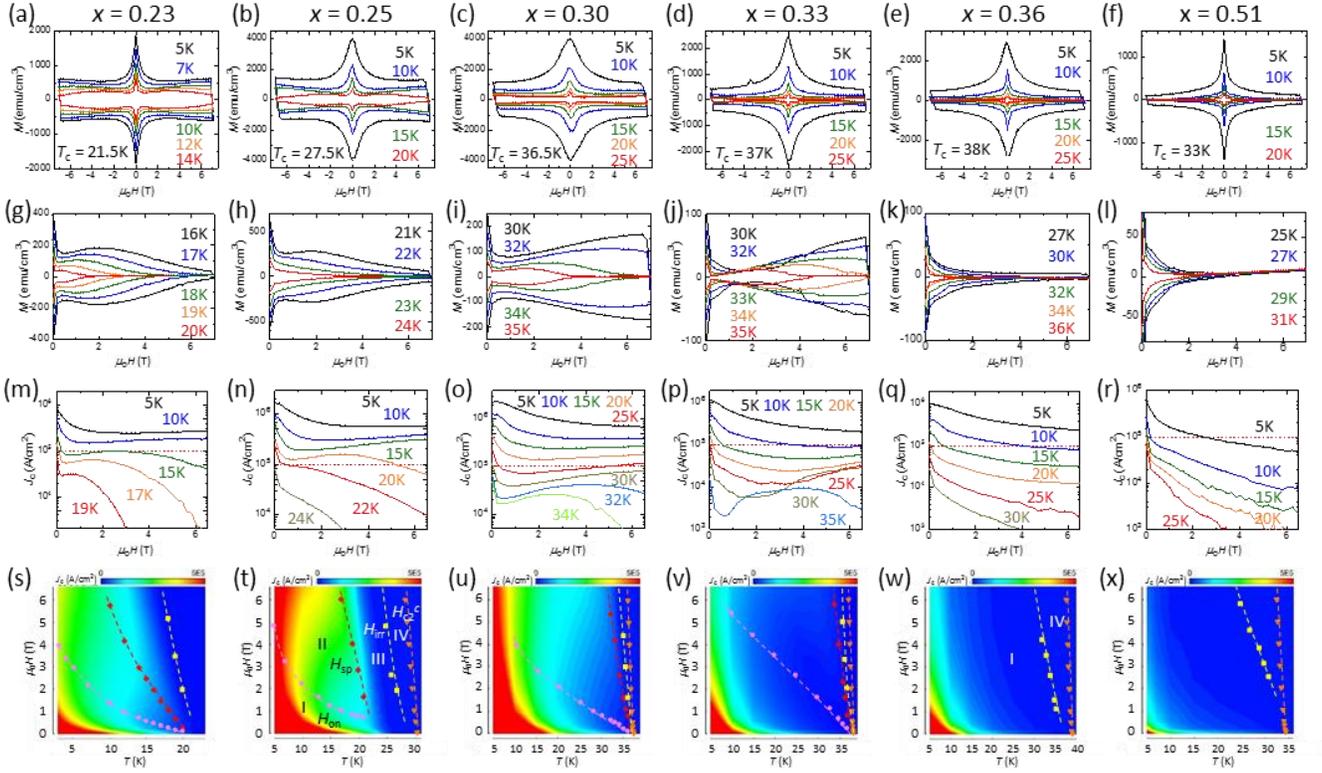

FIG. 2. (Color online) Magnetization hysteresis loops for samples; $x$ = 0.23, 0.25, 0.30, 0.33, 0.36, and 0.51 at $T \leq 0.7T_c$ (a)-(f) and $T \geq 0.7T_c$ (g)-(l). (m)-(r) Magnetic field dependence of critical current density for six samples. (s)-(x) Vortex phase diagram in forms of contour plots. Red (blue) color indicates high (low) $J_c$ region.

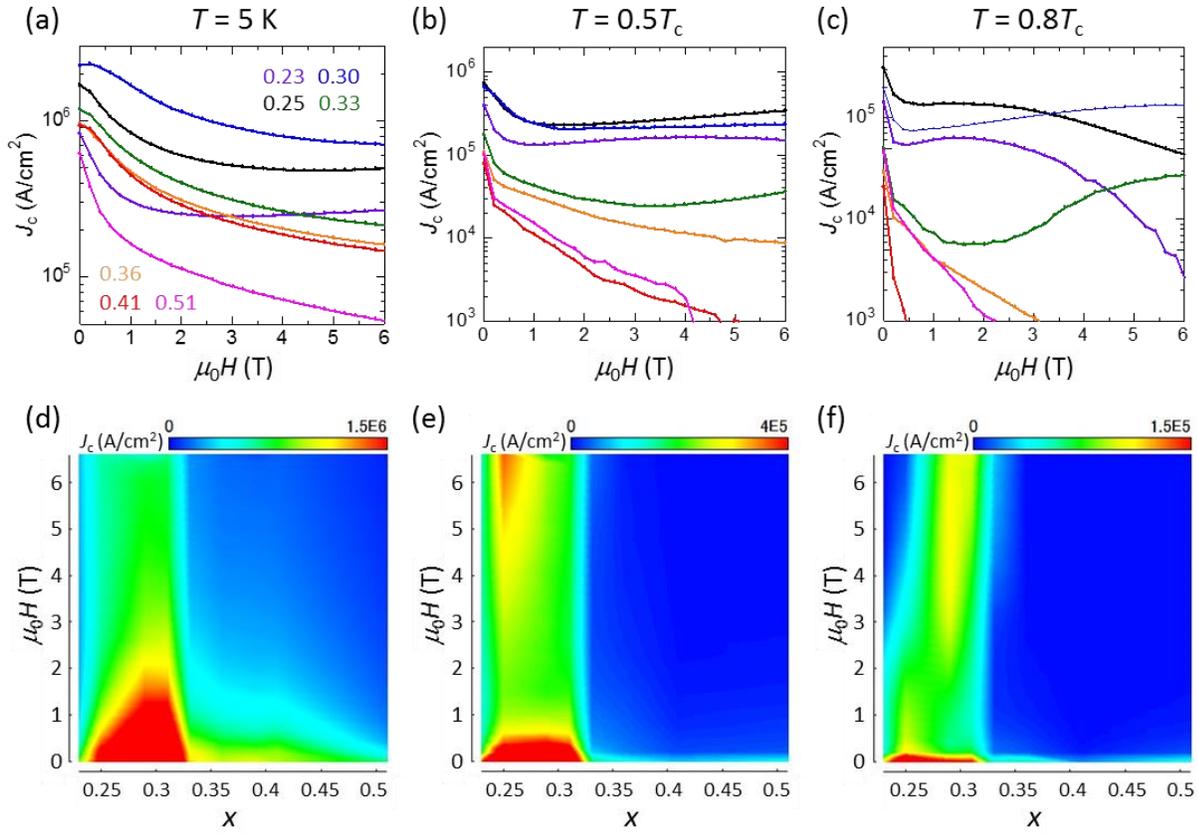

FIG. 3. (Color online) (a)-(c) Magnetic field dependence of critical current density for six samples at $T = 5$ K, $0.5T_c$, and $0.8T_c$. (d)-(F) Doping and field dependence of $J_c$ derived from the upper panels.

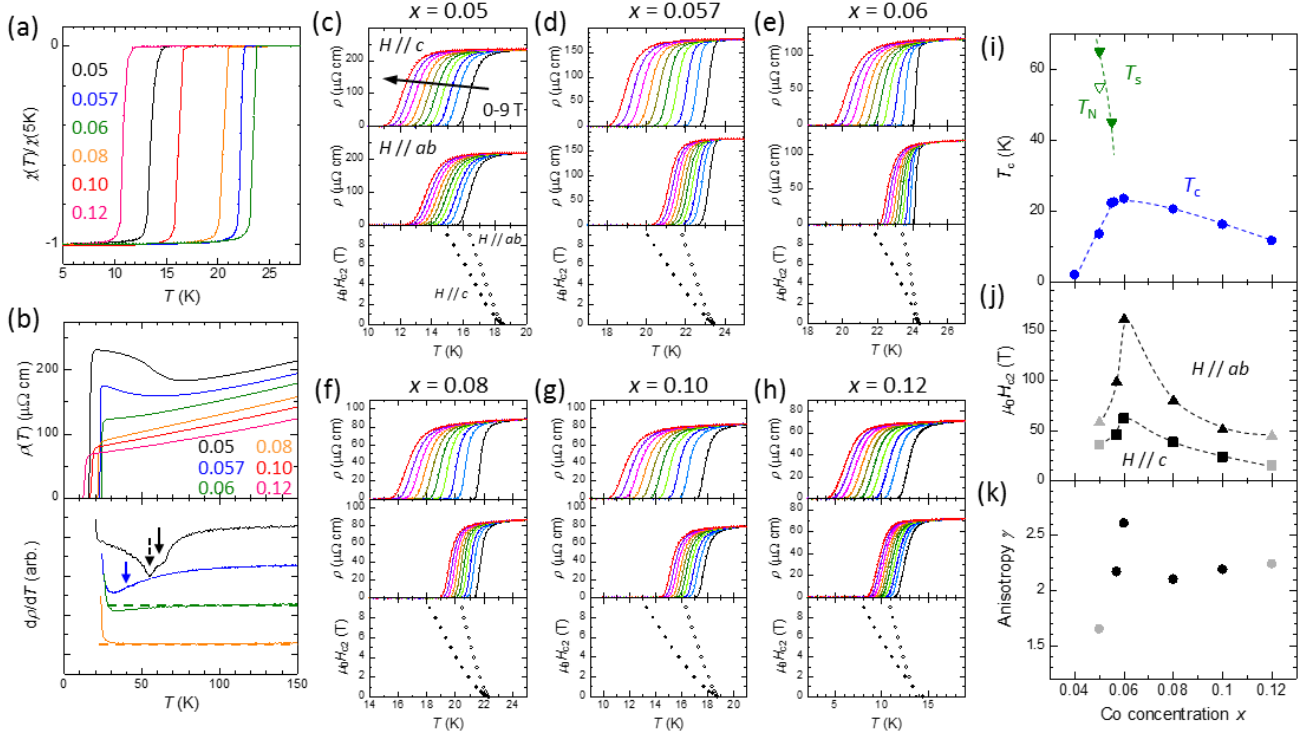

FIG. 4. (Color online) Characterization of Ba(Fe$_{1-x}$Co$_x$)$_2$As$_2$ single crystals with $x$ = 0.05-0.12. (a) Temperature dependence of magnetization. (b) Temperature dependence of in-plane resistivity (upper panel) and the derivative (lower panel). (c)-(h) In-plane resistivity in magnetic fields parallel to $c$ axis (upper panel) and $ab$ plane (lower panel). The bottom panel show $H_{c2}$. (i)-(k) Doping dependence of $T_c$ and $T_{s/N}$, $H_{c2}(0)$ with $H // c$ (filled circles) and $ab$ (open circles), and anisotropy $\gamma$ of $H_{c2}$ ((d$H_{c2}^{ab}$/d$T$)/(d$H_{c2}^{c}$/d$T$)).

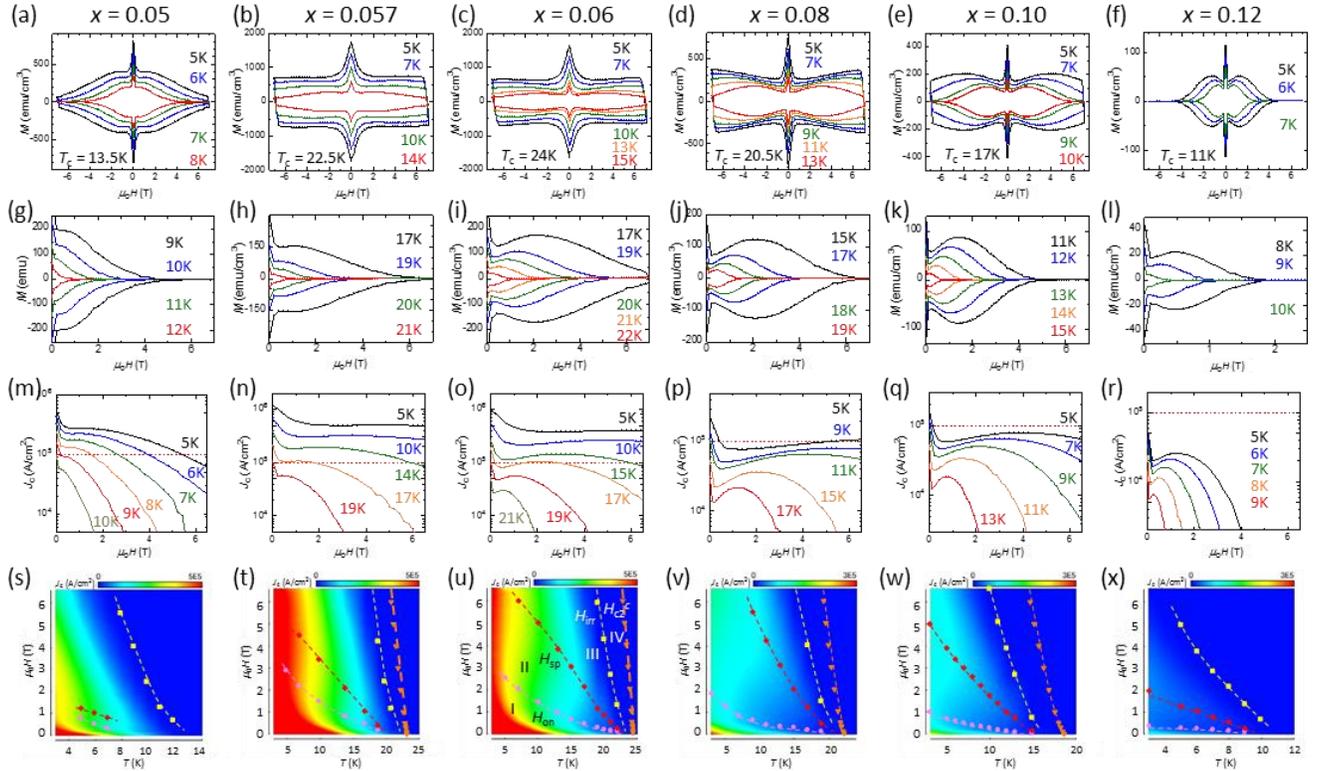

FIG. 5. (Color online) Magnetization hysteresis loops for samples with $x$ = 0.05, 0.057, 0.06, 0.08, 0.10, and 0.12 at temperatures $T \leq 0.7T_c$ (a)-(f) and $T \geq 0.7T_c$ (g)-(l). (m)-(r) Magnetic field dependence of critical current density for six samples. (s)-(x) Vortex phase diagrams. Note that color scale is changed from $x \leq 0.06$ to $x \geq 0.08$.

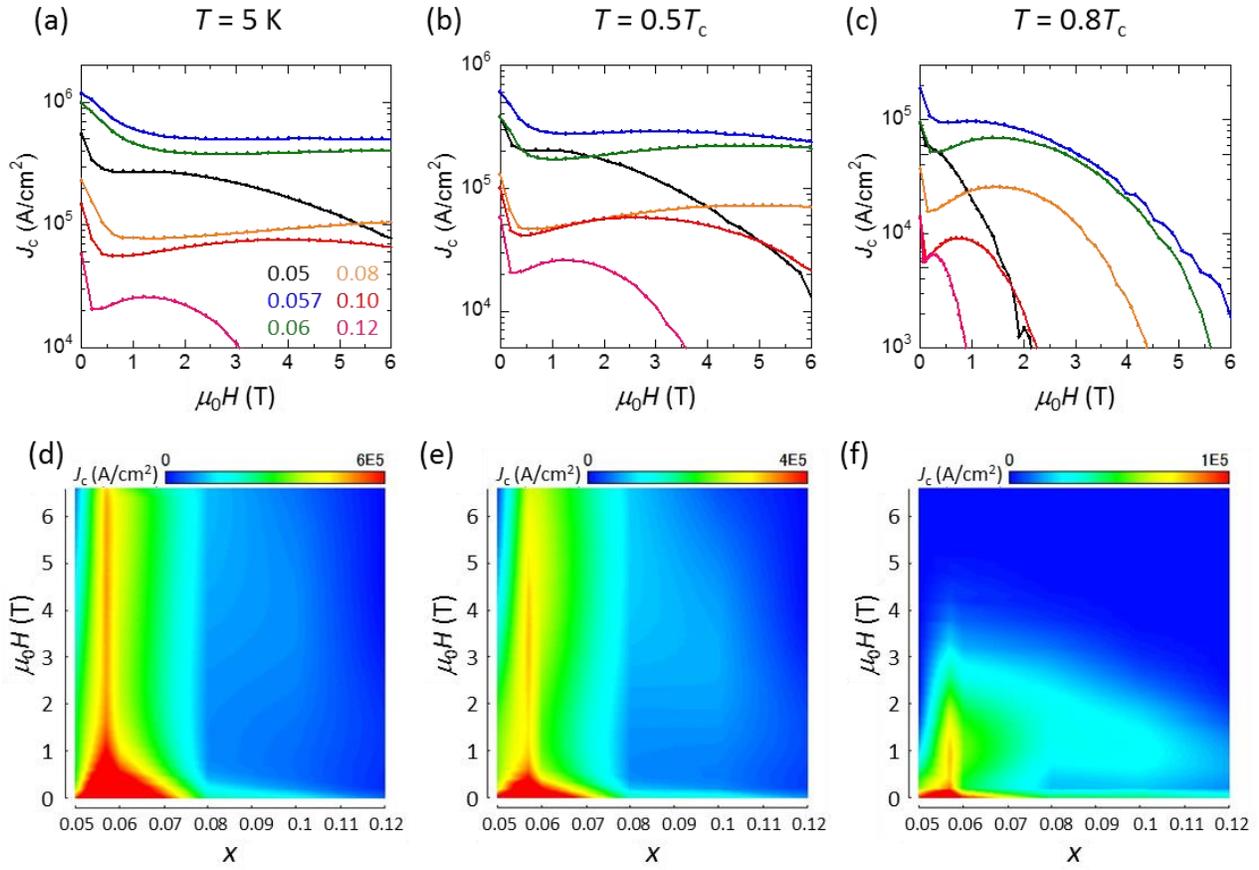

FIG. 6. (Color online) (a)-(c) Magnetic field dependence of critical current density for six samples at $T = 5$ K, $0.5T_c$, and $0.8T_c$. (d)-(f) Doping and field dependence of $J_c$ derived from the upper panels.

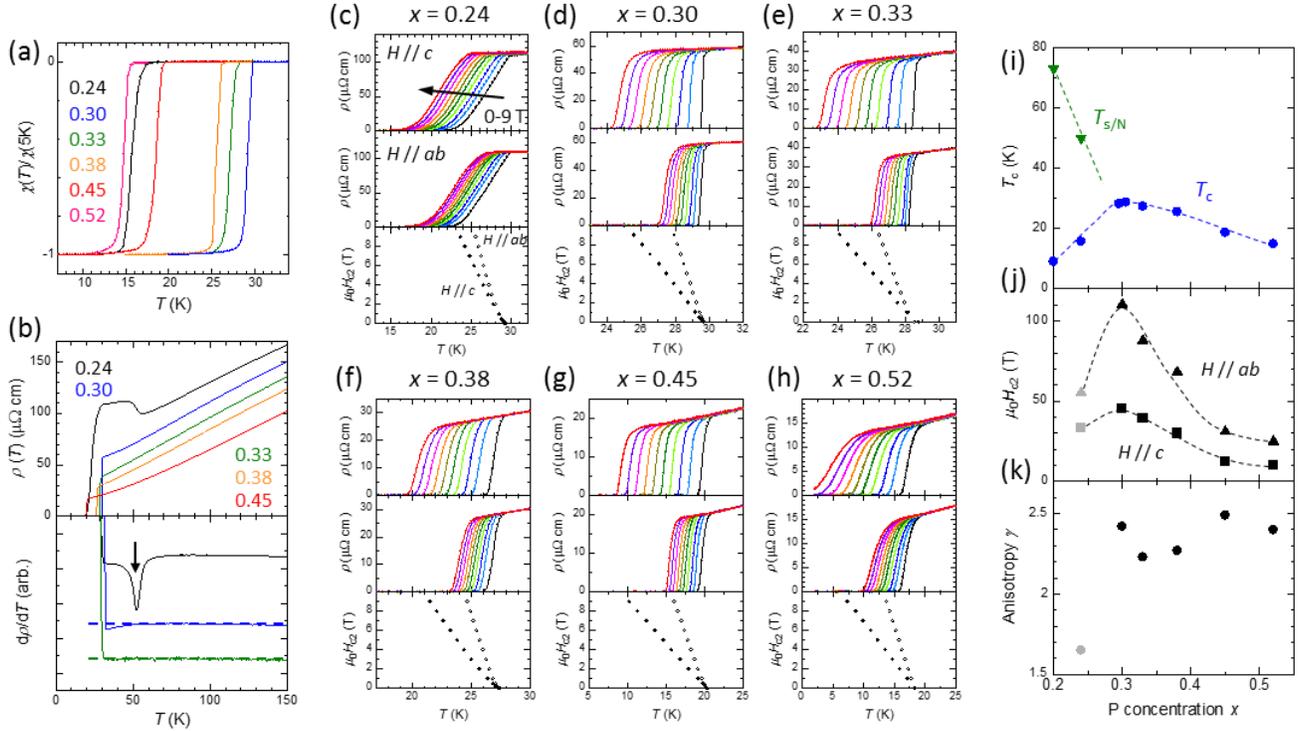

FIG. 7. (Color online) Characterization of BaFe$_2$(As$_{1-x}$P$_x$)$_2$ single crystals with $x$ = 0.24-0.52. (a) Temperature dependence of magnetization. (b) Temperature dependence of in-plane resistivity (upper panel) and the derivative (lower panel). (c)-(h) In-plane resistivity in magnetic fields parallel to $c$ axis (upper panel) and $ab$ plane (lower panel). The bottom panel show $H_{c2}$. (i)-(k) Doping dependence of $T_c$ and $T_{s/N}$, $H_{c2}(0)$ with $H // c$ (filled circles) and $ab$ (open circles), and anisotropy $\gamma$ of $H_{c2}$ (($dH_{c2}^{ab}/dT$)/($dH_{c2}^c/dT$)).

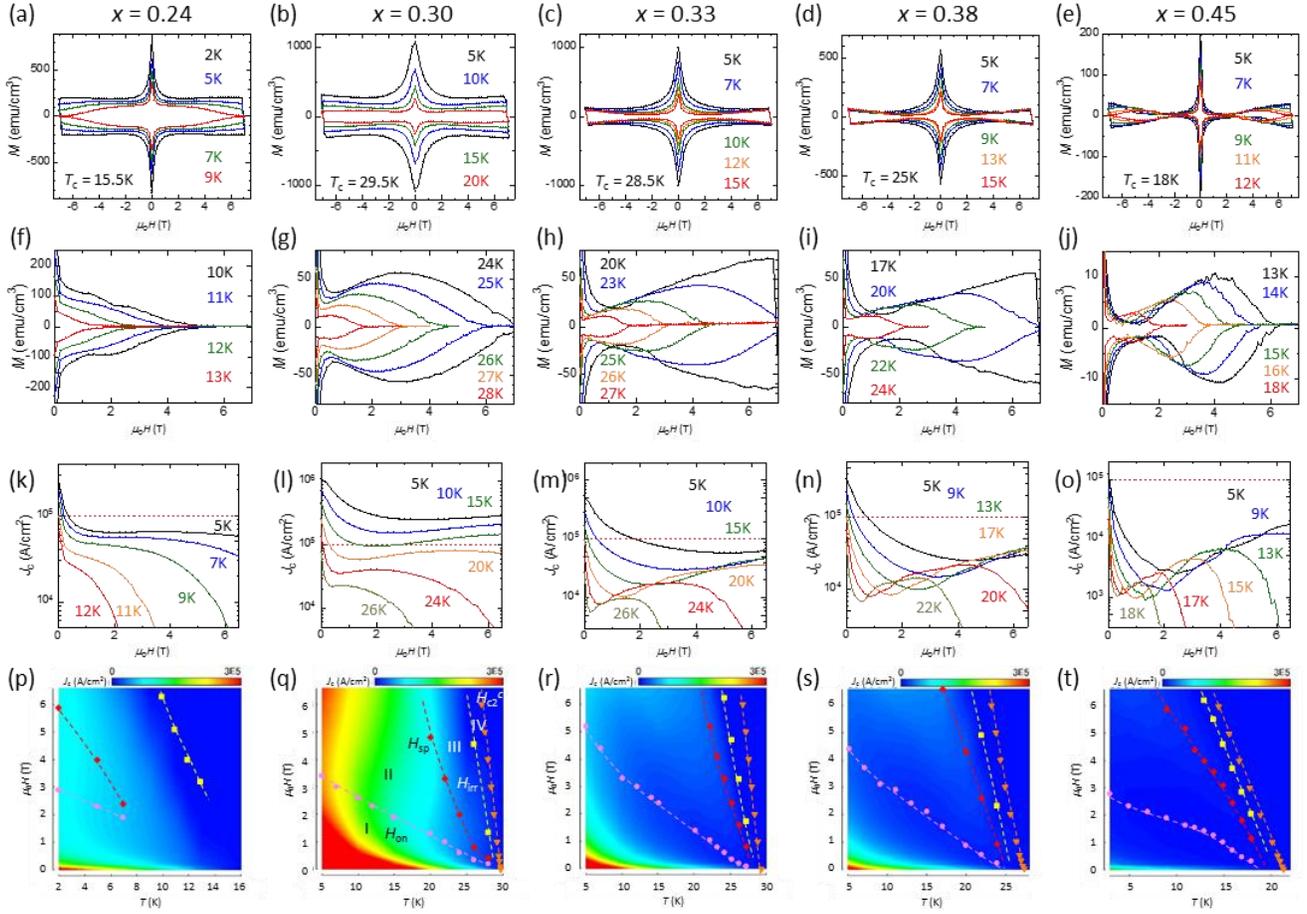

FIG. 8. (Color online) Magnetization hysteresis loops for six samples; $x = 0.24$, 0.30, 0.33, 0.38, and 0.45 at $T \leq 0.7T_c$ (a)-(e) and $T \geq 0.7T_c$ (f)-(j). (k)-(o) Magnetic field dependence of critical current density for six samples. (p)-(t) Vortex phase diagrams.

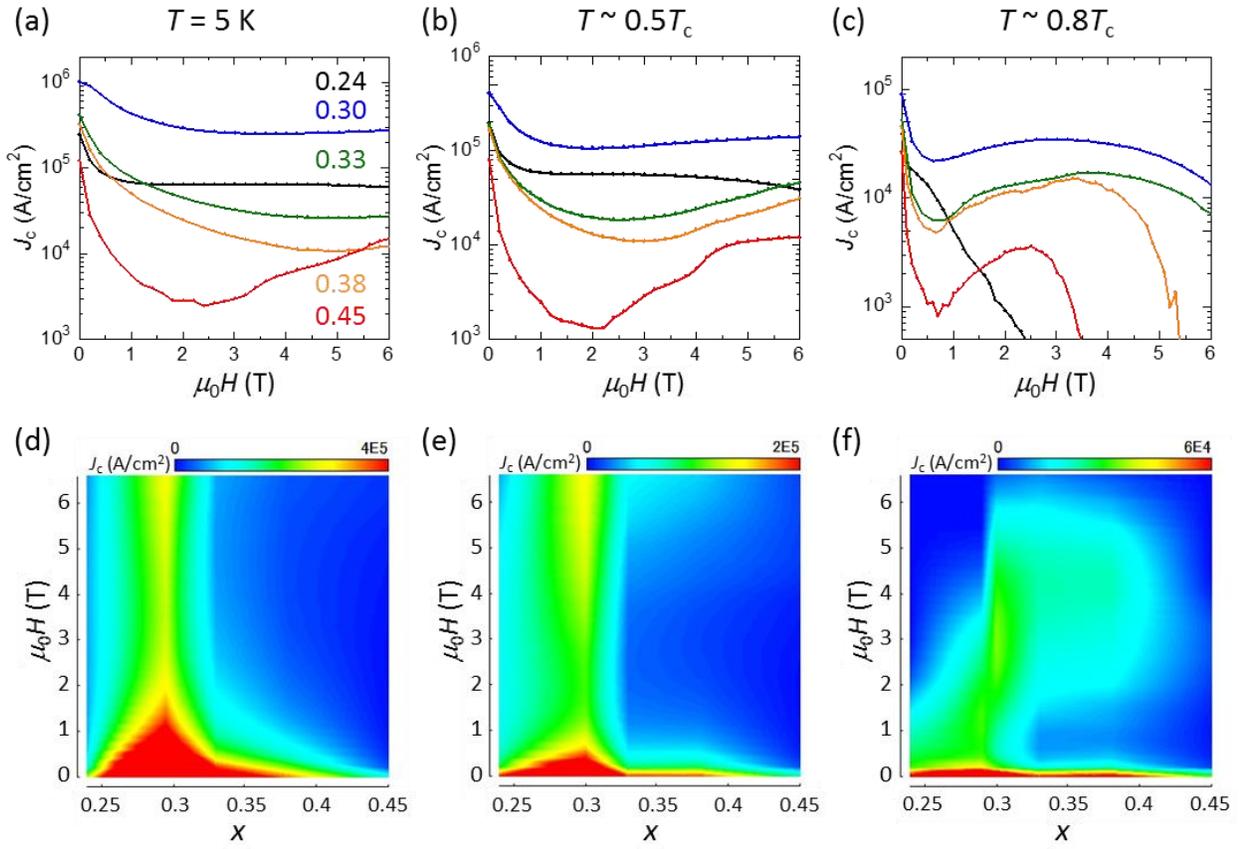

FIG. 9. (Color online) (a)-(c) Magnetic field dependence of critical current density at $T = 5$ K, $0.5T_c$, and $0.8T_c$. (d)-(f) Doping and field dependence of $J_c$ derived from the upper panels.

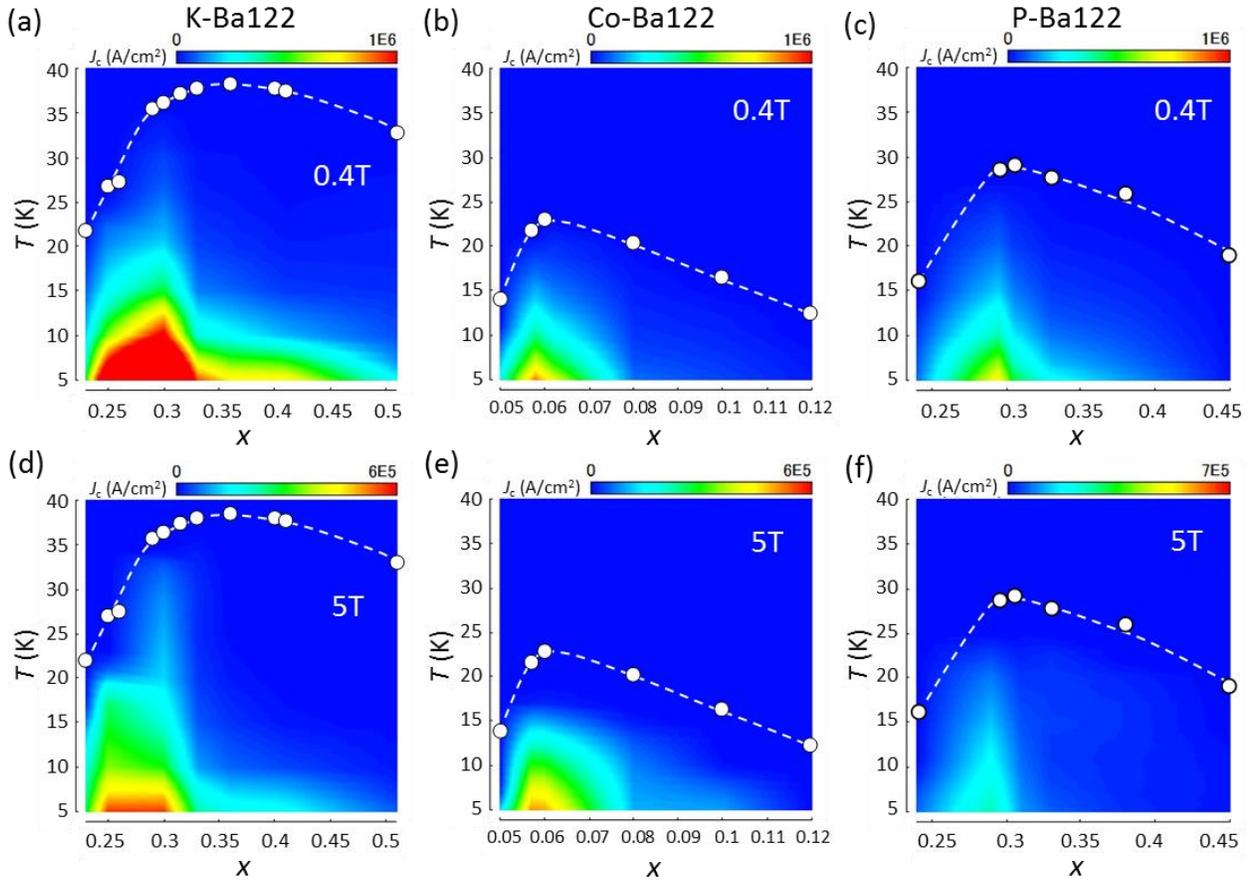

FIG. 10. (Color online) Doping and temperature dependence of critical current density (measured at $\mu_0H = 0.4$ T) for Ba$_{1-x}$K$_x$Fe$_2$As$_2$ (a), Ba(Fe$_{1-x}$Co$_x$)$_2$As$_2$ (b), and BaFe$_2$(As$_{1-x}$P$_x$)$_2$ (c). $T_c - x$ curves are also plotted in each panel.

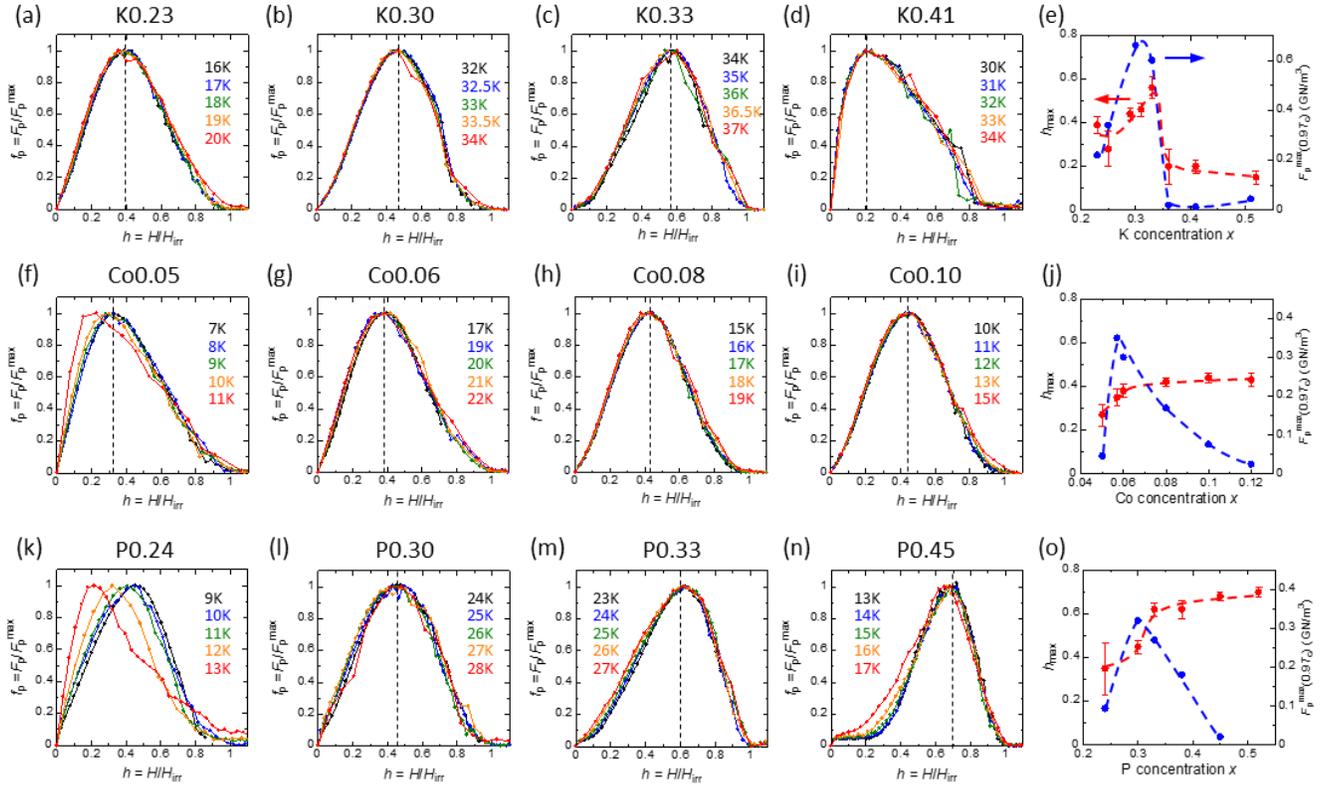

FIG. 11. (Color online) Normalized pinning force density ($f_p = F_p/F_p^{max}$) plotted against the reduced magnetic field ($h = H/H_{irr}$) for K-Ba122 (a)-(d), Co-Ba122 (f)-(i), and P-Ba122 (k)-(n). $x$ dependences of $h_{max}$ and $F_p^{max}(0.9T_c)$ for K-Ba122 (e), Co-Ba122 (j), and P-Ba122 (o).

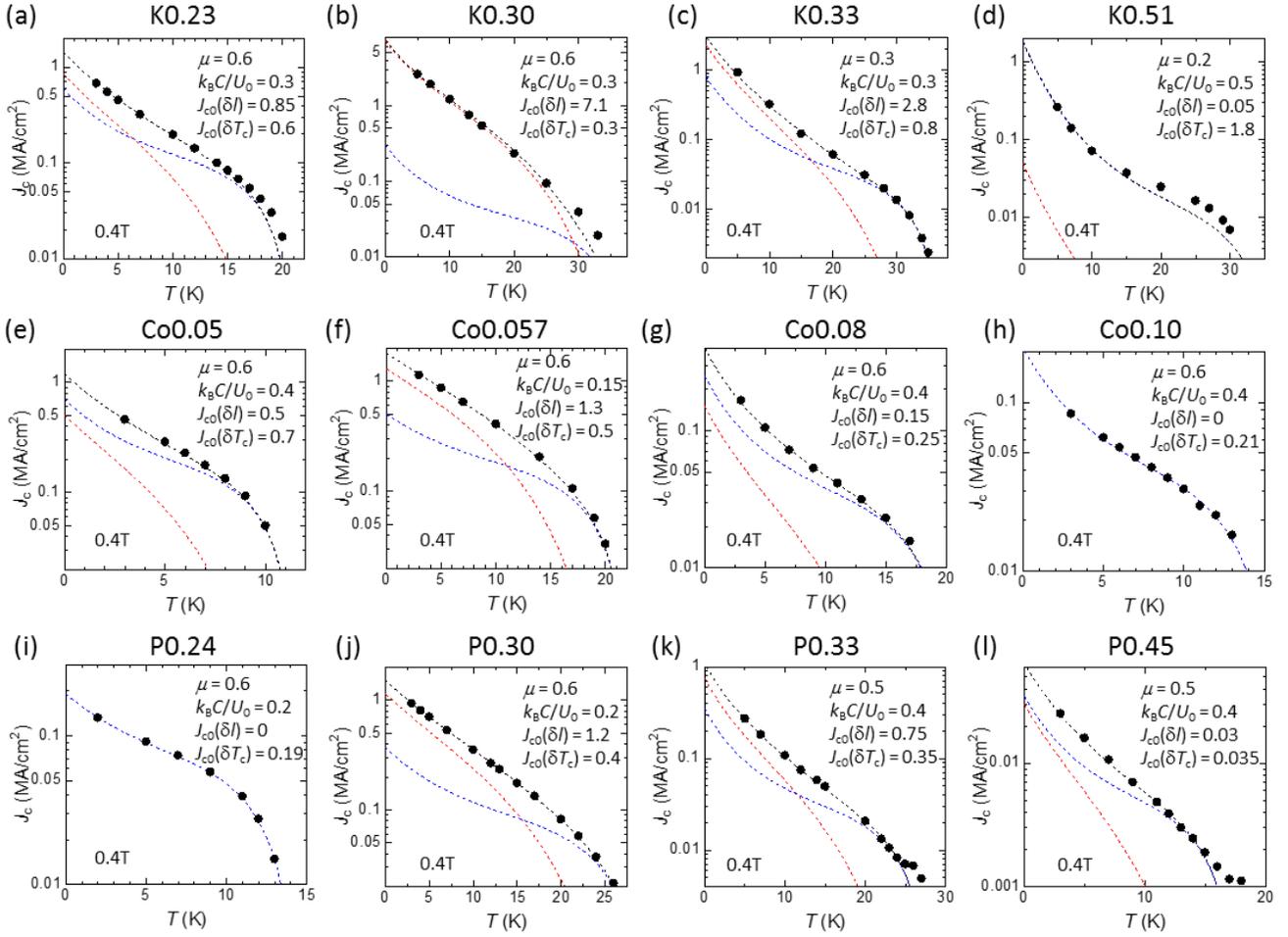

FIG. 12. (Color online) Temperature dependence of $J_c$ for selected compounds; $Ba_{1-x}K_xFe_2As_2$ ($x$ = 0.23, 0.30, 0.33, and 0.51) (a)-(d), $Ba(Fe_{1-x}Co_x)_2As_2$ ($x$ = 0.05, 0.057, 0.08, and 0.10) (e)-(h), and $BaFe_2(As_{1-x}P_x)_2$ ($x$ = 0.24, 0.30, 0.33, and 0.45) (i)-(l). The black-, blue-, and red-dashed curves indicate the total $J_c$, $\delta T_c$ pinning, and $\delta l$ pinning components, respectively. The parameters used for the fitting are shown in each panel (units for $k_BC/U_0$ and $J_{c0}$ are K$^{-1}$ and MA/cm$^2$, respectively).

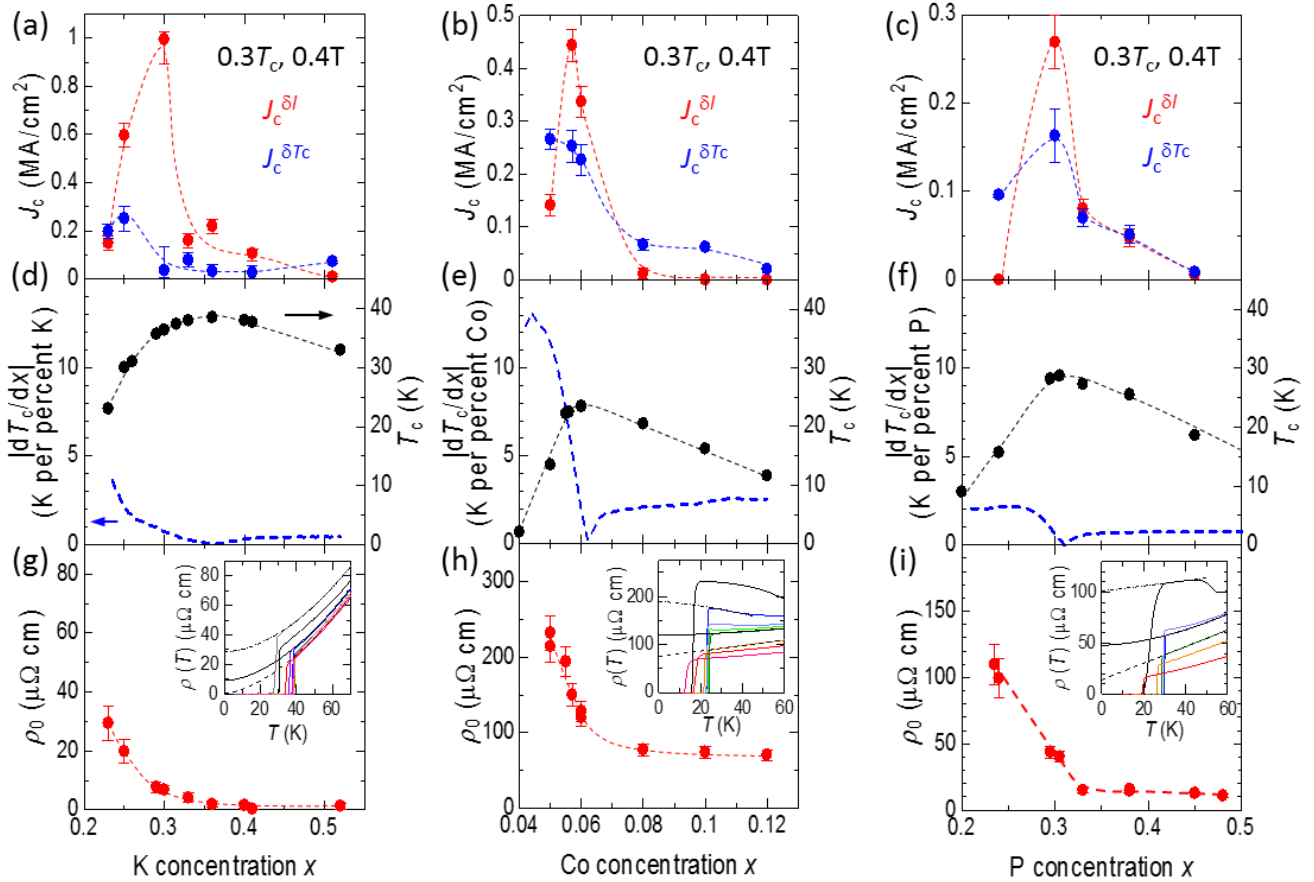

FIG. 13. (Color online) Doping dependence of $J_c^{\delta T_c}$ and $J_c^{\delta l}$ at $\mu_0 H = 0.4$ T and $T = 0.3 T_c$ (a)-(c), $T_c$ and magnitude of $dT_c/dx$ (d)-(f), and residual resistivity extracted from power-law fitting (several examples are shown in inset) (g)-(i) for K-Ba122 (left), Co-Ba122 (center), and P-Ba122 (right), respectively. The dashed lines are guides to the eyes.

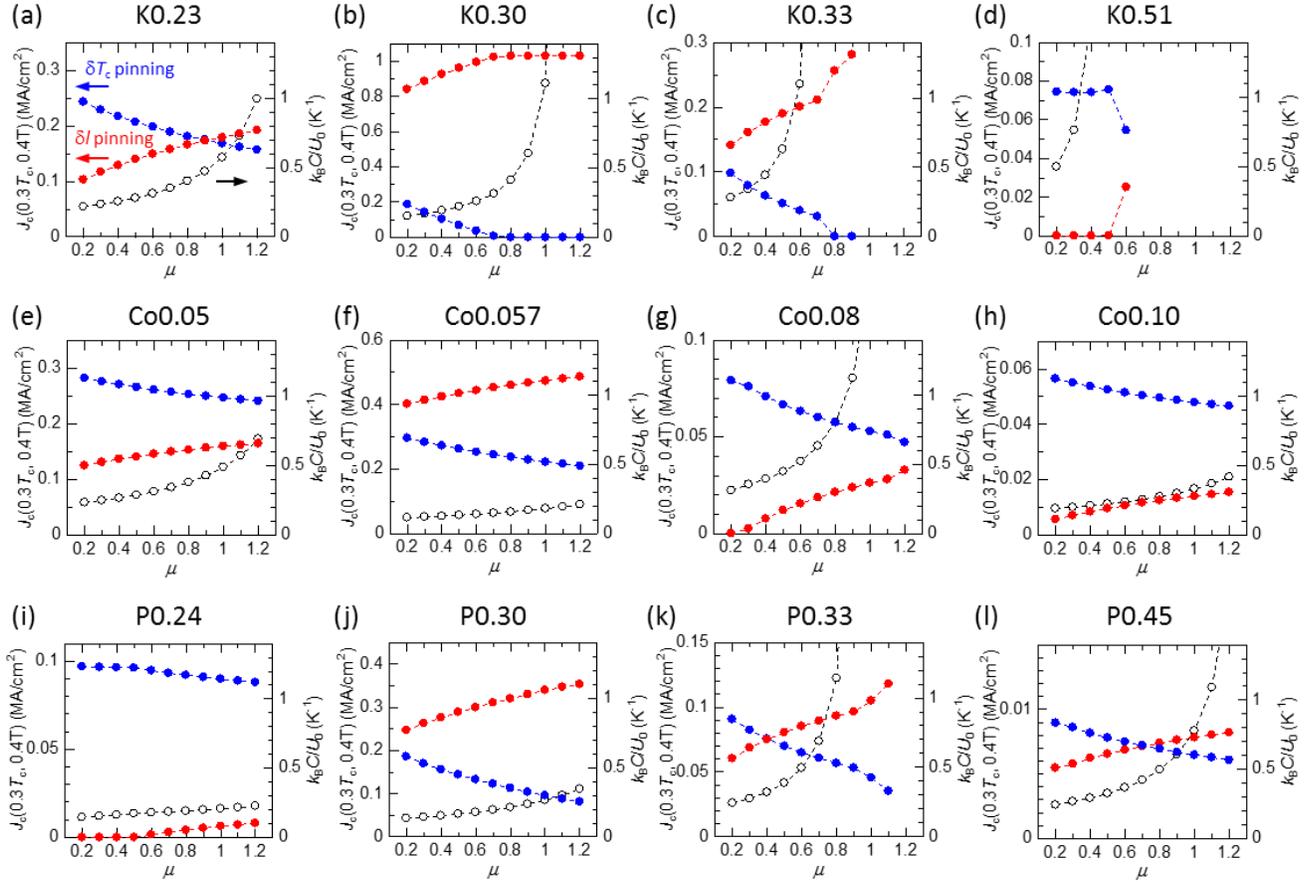

FIG. 14. (Color online) Dependence of the $\delta T_c$ pinning (blue circles) and $\delta l$ pinning (red circles) contributions on $\mu$ for the representative samples corresponding to Fig. 12. The derived $k_B C/U_0$ values are also plotted (black open circles).